\documentclass[fleqn,usenatbib]{mnras}

\usepackage{newtxtext,newtxmath}
\usepackage[T1]{fontenc}

\def\msun{M$_{\odot}$}
\def\Msun{M$_{\odot}$ }
\def\be{\begin{equation}}
\def\ee{\end{equation}}
\def\bi{\begin{itemize}}
\def\i{\item}
\def\ei{\end{itemize}}
\def\ben{\begin{enumerate}}
\def\een{\end{enumerate}}
\def\bea{\begin{eqnarray}}
\def\eea{\end{eqnarray}}
\def\bt{\begin{tabbing}}
\def\et{\end{tabbing}}
\def\gcc{gcm$^{-3}$}

\def\edo{\end{document}}

\def\M4{M$_4$}
\def\M6{M$_6$}
\def\lcm6{LCM$_6$}
\def\qcm6{QCM$_6$}

\def\wh3{W$_{\rm H,3}$}
\def\wh4{W$_{\rm H,4}$}
\def\wh5{W$_{\rm H,5}$}
\def\wh6{W$_{\rm H,6}$}
\def\wh7{W$_{\rm H,7}$}

\newcommand{\SpB}{\texttt{SPHINCS\_BSSN}\; }
\newcommand{\spB}{\texttt{SPHINCS\_BSSN}}

\newcommand{\Fu}{\texttt{FUKA}\;}
\newcommand{\fu}{\texttt{FUKA}}

\DeclareRobustCommand{\VAN}[3]{#2}
\let\VANthebibliography\thebibliography
\def\thebibliography{\DeclareRobustCommand{\VAN}[3]{##3}\VANthebibliography}

\usepackage{graphicx}	% Including figure files
\usepackage{amsmath}	% Advanced maths commands
\title[Fast ejecta in NSM]{Fast dynamic ejecta in neutron star mergers}

\author[S. Rosswog et al.]{
Stephan Rosswog,$^{1,2}$\thanks{E-mail: stephan.rosswog@astro.su.se}
Nikhil Sarin,$^{3,4}$
Ehud Nakar$^{5}$
and Peter Diener$^{6,7}$
\\
$^{1}$Hamburger Sternwarte, University of Hamburg, Gojenbergsweg 112, 21029 Hamburg, Germany\\
$^{2}$Department of Astronomy and Oskar Klein Centre, Stockholm University, 10619 Stockholm, Sweden \\
$^3$ Nordita, Stockholm University and KTH Royal Institute of Technology Hannes Alfvéns väg 12, SE-106 91 Stockholm, Sweden \\
$^4$ The Oskar Klein Centre, Department of Physics, AlbaNova, Stockholm University, SE-106 91 Stockholm, Sweden\\
$^5$ School of Physics \& Astronomy, Tel Aviv University, Tel Aviv 69978, Israel\\
$^6$ Center for Computation \& Technology, Louisiana State University, 70803, Baton Rouge, LA, USA\\
$^7$ Department of Physics \& Astronomy, Louisiana State University, 70803, Baton Rouge, LA, USA
}

\date{Accepted XXX. Received YYY; in original form ZZZ}

\pubyear{2022}

\begin{document}
\label{firstpage}
\pagerange{\pageref{firstpage}--\pageref{lastpage}}
\maketitle

\begin{abstract}
The ejection of neutron-rich matter is one of the most important consequences of a neutron star merger. While the bulk of the  matter is ejected at fast, but non-relativistic velocities ($\sim0.2c$), a small amount of
mildly relativistic dynamic ejecta have been seen in a number of numerical
simulations. Such ejecta  can have
far reaching observational consequences ranging from the shock breakout burst of gamma-rays promptly after the merger, to an early ($\sim 1$ hour post-merger)
blue  kilonova precursor signal, to synchrotron emission years after the merger ("kilonova afterglow"). These all potentially carry the imprint of the binary system parameters and the equation of state.
By analyzing Lagrangian simulations in full General Relativity, performed with the
code \spB, we identify two ejection mechanisms for fast ejecta: i) about 30\% of the ejecta with {$v> 0.4c$} are "sprayed out" from the shear interface 
between the merging stars and
escape along the orbital plane and ii) the remaining $\sim$ 70\% of the fast ejecta result from the central object "bouncing back"
after strong, general-relativistic compression. This "bounce component" is ejected in a rather isotropic way and reaches larger velocities (by $\sim0.1c$) so that its faster parts can catch up with and shock slower parts of the spray ejecta. Even for a case that promptly collapses to a black hole, we find fast ejecta with similar properties to the non-collapsing case, while slower matter parts are swallowed by the forming black hole. We discuss observational implications
of these fast ejecta, including shock breakout and kilonova afterglow.
\end{abstract}

\begin{keywords}
gravitational waves -- hydrodynamics -- radio continuum: transients
\end{keywords}

%%%%%%%%%%%%%%%%%%%%%%%%%%%%%%%%%%%%%%%%%%%%%%%%%%

%%%%%%%%%%%%%%%%% BODY OF PAPER %%%%%%%%%%%%%%%%%%

\section{Introduction}
Among the arguably most important consequences of  binary neutron star (BNS) and
neutron star black hole (NSBH) mergers is the ejection of neutron-rich  matter into space. This matter
can experience r-process nucleosynthesis \citep{lattimer74,symbalisty82,eichler89,rosswog99,freiburghaus99b}
and the initially extremely neutron-rich  matter provides ideal conditions to
forge the "strong r-process" elements beyond nucleon numbers of  $A\approx 130$,
see \cite{cowan21} for an excellent recent review on the r-process. The radioactivity of the 
freshly synthesized r-process nuclei can power electromagnetic transients ("kilonovae"
or "macronovae") \citep{li98,kulkarni05,rosswog05a,metzger10b,metzger19a}, an 
example of which has been observed in exquisite detail in the aftermath of the first multi-messenger 
BNS detection GW170817/AT2017gfo \citep[e.g.][]{coulter17,nicholl17,arcavi17,kilpatrick17,drout17,evans17}.\\
BNSs eject matter through various channels, each of which has its own characteristic time scale, electron fraction distribution and mass. 
The relative strength of these channels may vary with binary parameters (e.g. mass ratio) and nuclear matter equation of state (EOS).
The ejecta can be broadly grouped into "dynamic ejecta" (launched on dynamical time scales of ms), "winds" 
(typically hundreds of ms) and "torus ejecta" (seconds). 
For an up-to-date summary of the understanding of these ejection channels and 
for links to the (ample) original literature, we refer to the recent review
of \cite{rosswog24a}. In the present paper we focus exclusively on
dynamic ejecta. While the average velocities of dynamic ejecta are $\sim 0.2c$, numerical simulations
\citep{hotokezaka13,bauswein13a,kyutoku14,kiuchi17,radice18a,hotokezaka18a,dean21,nedora21,rosswog22b,combi23} have found that a small fraction of the 
ejecta reaches mildly relativistic velocities at least up to $\sim 0.8c$ (but higher velocities likely exist in nature without being resolved in the simulations).\\
Such fast ejecta can have interesting observational 
implications. First, the breakout of the shock driven by the relativistic jet (launched following the merger) and its cocoon through the fast ejecta produces a 
burst of gamma rays, which is probably the source of the gamma-rays observed in GW 170817 \citep{Gottlieb2018,Nakar2020,gutierrez24}. Second, due 
to their homologous structure, the outermost ejecta are fastest and the velocities in the leading parts could be so large
that many of the neutrons that are initially present would not be captured by seed nuclei and instead escape as free and finally decaying neutrons.
If true, this is expected to produce a bright UV/optical counterpart 
on time scales of minutes to hours \citep{kulkarni05,metzger15a}.  Finally, such mildly relativistic matter can at later time cause a "kilonova afterglow" due to synchrotron 
emission \citep{nakar11a,mooley17,hotokezaka18a,hajela22,sadeh23,sadeh24}. While these small amounts of fast dynamical
ejecta are difficult to resolve numerically, the fact that several groups with different simulation methodologies 
find them, provides some confidence that these fast ejecta are real and not just a mere numerical artefact.\\
It is less clear, however, by which mechanism(s) these ejecta are actually produced. While \cite{metzger15a}
state, based on the simulations of \cite{bauswein13a}, that this matter "originates from the shock-heated 
interface between the NS's", \cite{radice18a}
note that  "most of the fast moving ejecta instead appear to originate when the shock launched from
the first bounce of the remnant breaks out of the forming ejecta cloud". While these authors explore a
large number of different cases (and do find substantial differences depending on mass ratio and equation of state;
EOS hereafter), it is nontrivial to trace the origin and history of the fast ejecta component
with their Eulerian approach.
\cite{dean21} explored how  the properties of the fast ejecta depend on the numerical resolution, but their
high resolution comes at the price of having to approximate the merger as an axisymmetric collision 
with Newtonian self-gravity. Clearly, if the fast ejecta should be dominated by the "bounce mechanism" seen
by \cite{radice18a}, then such approximations have no chance to achieve a quantitatively correct results
because the re-bounce in Newtonian gravity is {\em much} less violent than in General Relativity (GR).\\
Here we study the fast ejecta with a methodology that is different from the above described investigations.
First, while we use Lagrangian particles as \cite{bauswein13a}/\cite{metzger15a}, our hydrodynamic
simulations are performed in fully dynamical GR (contrary to their use of an approximate, conformally flat metric) and we 
employ roughly an order of magnitude more particles. Second, compared to the Eulerian studies 
\citep{radice18a,dean21}, it is comparatively easy in our approach to trace the history of 
ejected material backwards (provided that the interesting matter can be identified in the first 
place)\footnote{For Eulerian studies with a focus on tracking matter via tracer particles, see e.g. \cite{zenati23,zenati24}.}.
So far, however, we have only run a restricted set of binary mergers 
which use piecewise polytropic approximations to nuclear matter equations of state together with some 
thermal pressure component. Nevertheless, given the unique combination of full GR and the ability 
to accurately track matter history, we consider it instructive to explore the ejection mechanism and 
possible observational consequences even for a restricted set of simulations.\\
Our paper is structured as follows. In Sec.~\ref{sec:sim_sum} we summarize the performed simulations, in Sec.~\ref{sec:morphology} 
we describe the morphology of the different classes of merger outcomes, in Sec.~\ref{sec:mechanism} we analyze the mechanism(s) 
by which these fast ejecta are launched,  in Sec.~\ref{sec:properties} we analyze their properties and discuss observational implications 
before we
finally summarize our findings in Sec.~\ref{sec:summary}.

%%%%%%%%%%%%%%%%%%%%%%%%%%%%%%%%%%%%%%
\begin{figure*}
        \vspace*{-0.cm}
	 \centerline{
 \includegraphics[width=2.\columnwidth]{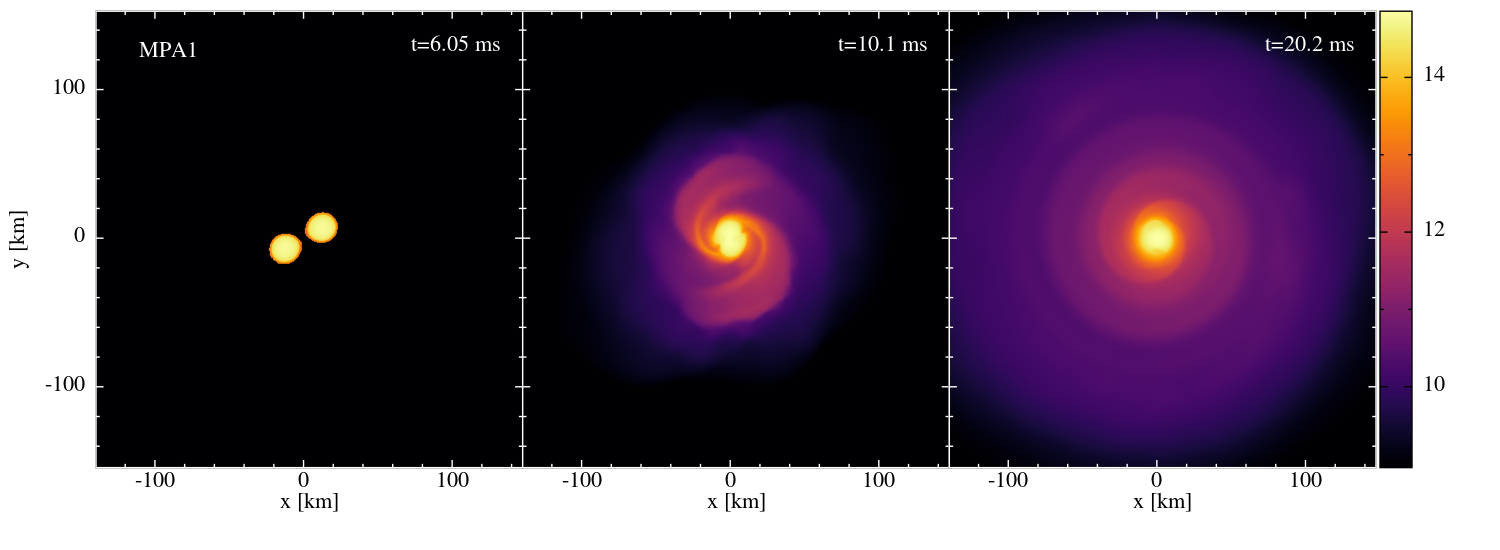} }
	 \vspace*{-0.3cm}		
         \caption{Logarithmic density distribution ($g/cm^3$) in the orbital plane of the simulation with $2 \times 1.3$ \Msun and the MPA1-EOS 
         (run \texttt{MPA1\_13\_13}). The color bar shows the logarithm of mass density in cgs units.}
    \label{fig:morph_MPA1_2x13}
\end{figure*}
%%%%%%%%%%%%%%%%%%%%%%%%%%%%%%%%%%%%%%

\section{Summary of the performed simulations}
\label{sec:sim_sum}
Our simulations are performed with the new, Lagrangian Numerical
Relativity code \SpB the methods of which have been laid out in substantial detail in a 
series of recent papers \citep{rosswog21a,diener22a,rosswog22b,rosswog23a,rosswog24c} and will not be repeated here. 
 Our code 
solves the full set of Einstein equations using the BSSN formalism \citep{shibata95,baumgarte99,alcubierre08,baumgarte10}
on an adaptive mesh. The matter evolution is performed with freely moving Lagrangian
particles that are evolved according to a modern version of Smooth Particle Hydrodynamics (SPH),
see \citet{monaghan05,rosswog09b,springel10a, price12a} and \cite{rosswog15c} for general reviews of the method.
\SpB has massively profited from our longer-term
development effort to enhance the accuracy of the SPH method
\citep{rosswog10a,rosswog10b,rosswog15b,rosswog20a,rosswog20b,rosswog21a,diener22a,rosswog22b,rosswog25a}.\\
% initial conditions
Our initial conditions for binary neutron stars are constructed in two steps. First, we need to find 
the spacetime and matter properties of a specified combination of binary parameters and EOS. For this step we use the library \Fu   \citep{papenfort21,kadath}.
In a second step, we need
to place SPH particles so that they accurately represent the solution found by \fu. To avoid numerical noise,
we want to use (close to) equal mass SPH particles that are placed in a way so that the SPH density 
estimate agrees as closely as possible with the \Fu solution. We achieve this via the "Artificial
Pressure Method" (APM), originally suggested in a Newtonian context \citep{rosswog20a}. The main
idea is to start with a guessed particle distribution, then measure the SPH-density value and to translate 
the difference between the measured SPH-density and the desired density profile into an "artificial pressure". 
The corresponding pressure gradient is used in a momentum-like equation to drive the particles into locations where they minimize their
density error. This method, extended to relativistic binary systems, has been implemented in the
code \texttt{SPHINCS$\_$ID} which is described in detail in \cite{diener22a} and further refined in \cite{rosswog23a}.\\
% simulations
We focus here on irrotational
binary systems and use four equations of state: SLy \citep{SLY_eos}, APR3 \citep{akmal98}, MPA1
\citep{MPA1_eos} and MS1b \citep{mueller96}. Based on their maximal Tolman-Oppenheimer-Volkoff masses and the tidal deformabilities of 1.4 \Msun neutron stars
(numbers taken from \cite{pacilio22}),
we consider SLy as (probably too) "soft" ($M_{\rm TOV}= 2.05$ \msun; $\Lambda_{1.4}= 297$), 
APR3 ($M_{\rm TOV}= 2.39$ \msun; $\Lambda_{1.4}= 390$)
and MPA1 ($M_{\rm TOV}= 2.46$ \msun; $\Lambda_{1.4}= 487$) as two rather realistic EOSs, while MS1b ($M_{\rm TOV}= 2.78$ \msun; $\Lambda_{1.4}= 1250$) is likely too stiff \citep{abbott17b}.
The cold parts of these nuclear EOSs are parametrized via piecewise polytropes 
as described in \cite{read09} and they are enhanced by a thermal pressure component \citep[e.g.,][]{shibata05b,roberts11,hotokezaka13} with adiabatic exponent
$\Gamma_{\rm th}=1.75$, see the appendix of \cite{rosswog22b} for implementation details. Each binary system
is modelled with 2 million SPH particles for the fluid and, for the spacetime evolution, we use initially seven mesh refinement levels, 
each with 193 grid points, and an outermost extension  in each coordinate direction of $\approx 2268$ km. The resolution on
the finest initial grid level is $\Delta x= 369$ m. As described in detail in \cite{rosswog23a}, our code automatically refines 
the mesh further in case this is needed, e.g. in a collapse to a black hole.  For the current code version, simulations with
substantially larger resolution are prohibitively slow, but this will be improved in the next  code revision.
Our simulations are summarized in  Tab.~\ref{tab:runs}, all of them start with an initial separation of 
$a_0=45$~km.
\begin{table}
	\centering
	\caption{Performed simulations. }
	\label{tab:runs}
	\begin{tabular}{clcccc} % four columns, alignment for each
		\hline
		label & EOS & M$_1$ & M$_2$ & $t_{\rm end}$  & remark\\
		         		 \hline
		\texttt{SLy\_13\_13} & SLy      & 1.30 & 1.30  & 28.3 &\\
        \texttt{SLy\_14\_14} & SLy      & 1.40 & 1.40  & 14.9 & prompt collapse\\
        \texttt{APR3\_13\_13} & APR3    & 1.30 & 1.30  & 35.4 \\
        \texttt{APR3\_14\_14} & APR3    & 1.40 & 1.40  & 33.0 \\
        \texttt{MPA1\_13\_13} & MPA1    & 1.30 & 1.30  & 37.5 \\
        \texttt{MPA1\_14\_14} & MPA1    & 1.40 & 1.40  & 27.2 \\
        \texttt{MPA1\_12\_15} & MPA1    & 1.20 & 1.50  & 17.7 \\
        \texttt{MPA1\_12\_18} & MPA1    & 1.20 & 1.80  & 12.8 \\
        \texttt{MS1b\_13\_13} & MS1b    & 1.30 & 1.30  & 31.4 \\
        \texttt{MS1b\_14\_14} & MS1b    & 1.40 & 1.40  & 30.1 \\
        \hline
	\end{tabular}
\end{table}
\section{Morphological overview}
\label{sec:morphology}
\begin{figure*}
    \centering
    \includegraphics[width=2.1\columnwidth]{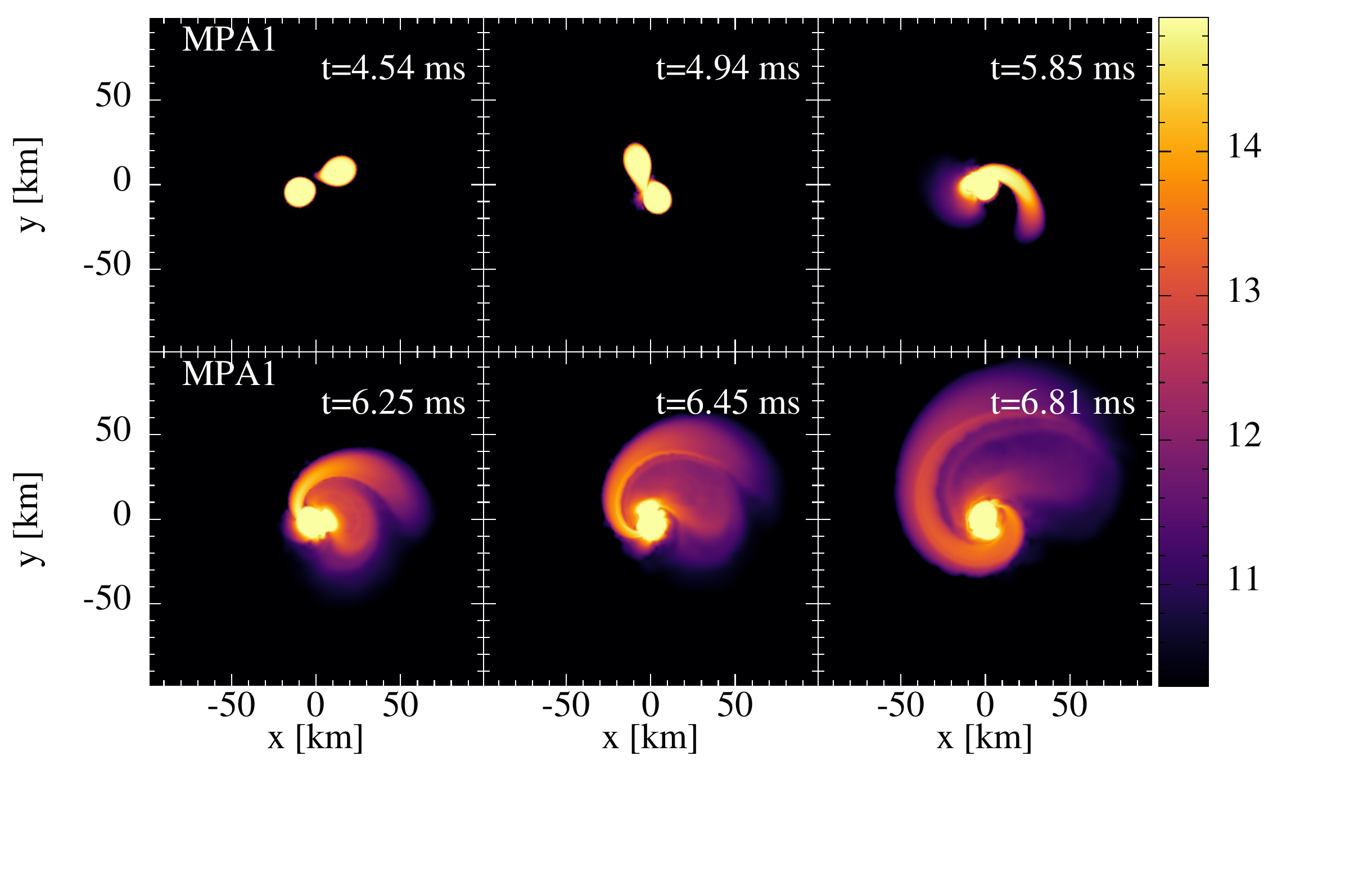}
    \vspace*{-2cm}	
    \caption{Density evolution for run \texttt{MPA1\_12\_18} (MPA1-EOS, masses of 1.2 and 1.8 \msun). The color bar shows the logarithm of mass density in cgs units.}
    \label{fig:morph_MPA1_12_18}
\end{figure*}

\begin{figure*}
    \centering
    \includegraphics[width=2.15\columnwidth]{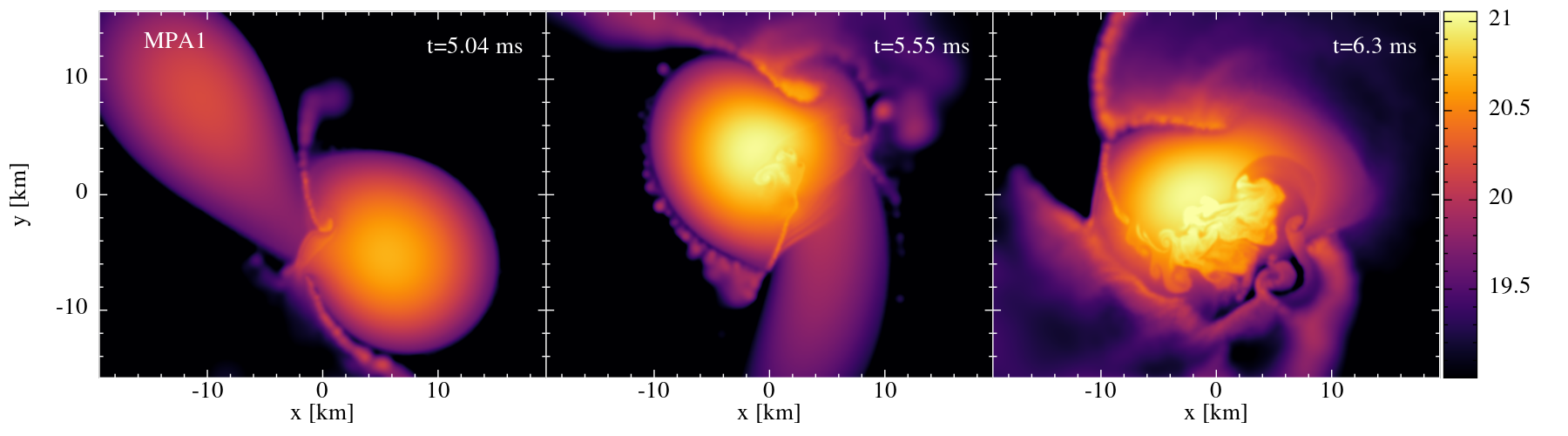}
    \vspace*{0cm}	
    \caption{Zoom into the impact region of run \texttt{MPA1\_12\_18} 
    (MPA1-EOS, masses of 1.2 and 1.8 \msun) which is also shown in the previous plot. Color-coded is the logarithm of the specific internal energy (cgs units).}
    \label{fig:impact_12_18}
\end{figure*}

\begin{figure*}
        \vspace*{-0.cm}
	 \centerline{
	\includegraphics[width=2.1\columnwidth]{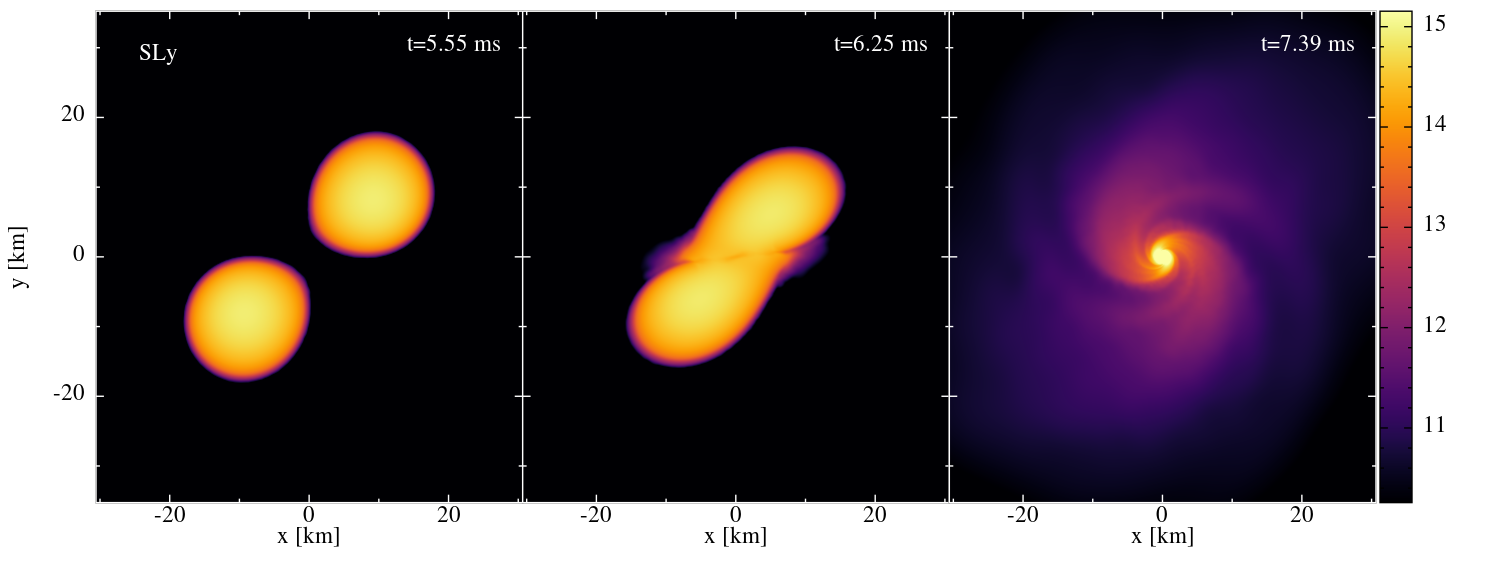}	}
	 \vspace*{-0.3cm}		
         \caption{Density in orbital plane for the case where the merger
                  results in a prompt black hole formation, run \texttt{SLy\_14\_14}(2 $\times$1.4 \msun; SLY-EOS).
                  The color bar shows the logarithm of mass density in cgs units.} 
    \label{fig:prompt_collapse}
\end{figure*}

\begin{figure*}
        \vspace*{-0.cm}
	 \centerline{
	\includegraphics[width=2.1\columnwidth]{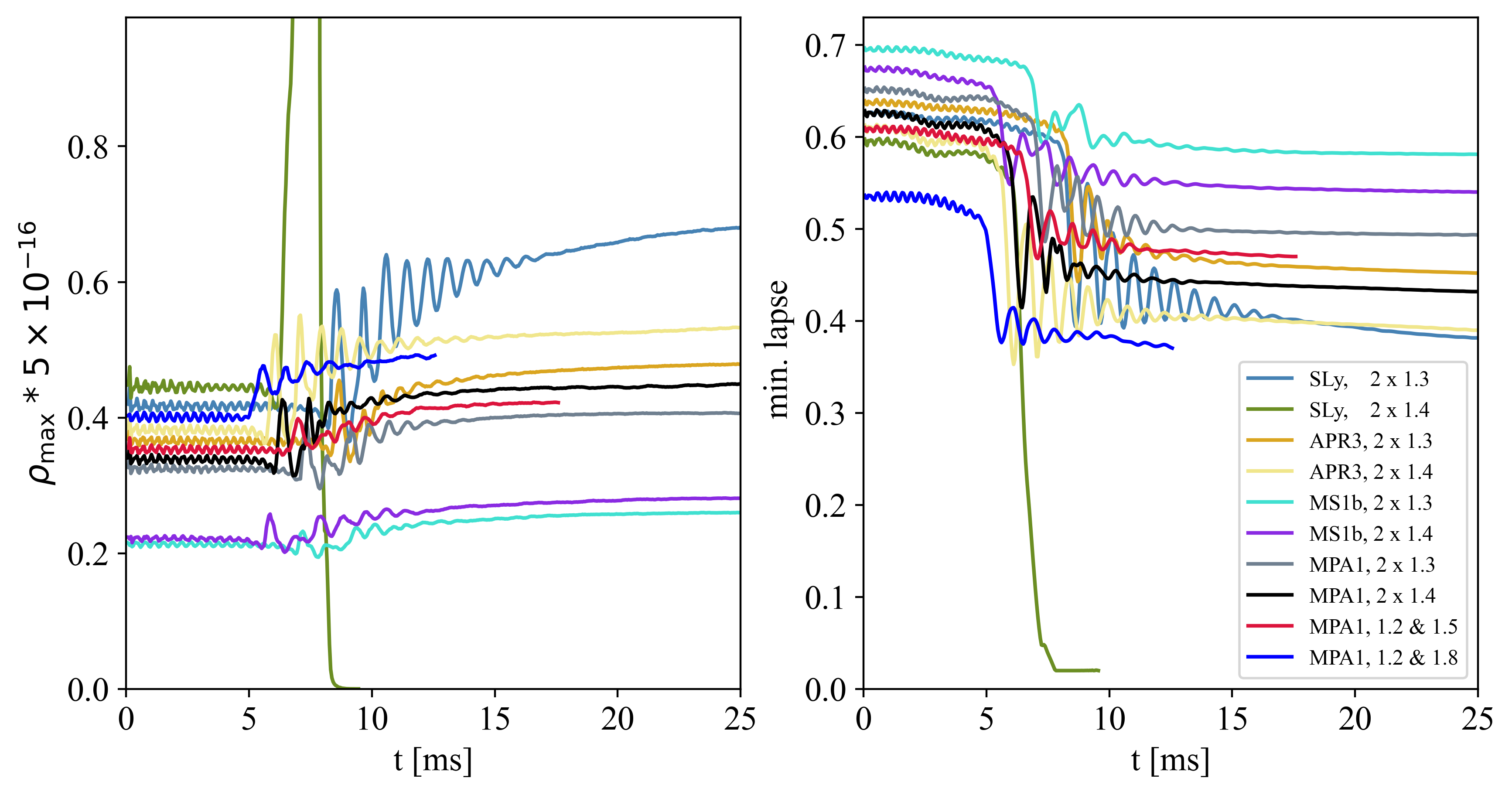}	}
	 \vspace*{-0.3cm}		
         \caption{Maximum values (in cgs, scaled) of the mass density (left) and minimum values of the lapse function (right), both at the particle positions.} 
    \label{fig:rho_lapse}
\end{figure*}

We first briefly illustrate the different classes of outcomes. 
We show in Fig.~\ref{fig:morph_MPA1_2x13} the density distribution in the orbital plane of run \texttt{MPA1\_13\_13}, representative for equal mass cases
where a central remnant survives. Here and throughout the paper, the run labels correspond to the equation of state  followed by the component masses. After the remnant has settled into a stationary state, the 
central remnant keeps shedding mass in the form of spiral waves.  Fig.~\ref{fig:morph_MPA1_12_18} 
shows our most extreme mass ratio simulation with $q= 2/3$, \texttt{MPA1\_12\_18}. As expected, this encounter results 
in one dominant tidal tail, mostly formed from the light neutron star \citep{rosswog00,korobkin12a,rosswog13b,dietrich17b,papenfort22}. It is worth noting that the heavier 
star suffers a severe impact with strong shocks and serious inner disturbance, see
Fig.~\ref{fig:impact_12_18}, where we show the specific internal energy in the collision region. The shear-induced instabilities are clearly visible in panels 
two and three, these regions should lead to a strong and very fast amplification of the magnetic
field strength, see e.g. \cite{price06,kiuchi15,palenzuela22,aguilera24,kiuchi24}. To illustrate the last class, the direct collapse
to a black hole, we show in Fig.~\ref{fig:prompt_collapse} the density evolution of run \texttt{SLy\_14\_14}. In this case, the central object begins
to contract immediately (see also the green curve in Fig.~\ref{fig:rho_lapse}) until a horizon forms. In the simulation, we remove
particles --very conservatively-- once their lapse functions become smaller
than $\alpha_{\rm cut}=0.02$. At this point the particles are already safely
inside of the forming apparent horizon \citep{rosswog23a} and their removal
does not have an impact on the outside evolution. \\
%%%%%%%%%%%%%%%%%%%%%%%%%%%%%%%%%%%%%
%%%%%%%%%%%%%%%%%%%%%%%%%%%%%%%%%%%%%%
%%%%%%%%%%%%%%%%%%%%%%%%%%%%%%%%%%%%%%
%%%%%%%%%%%%%%%%%%%%%%%%%%%%%%%%%%%%%%
In Fig.~\ref{fig:rho_lapse} we show the maximum mass densities and 
the minimum values of the lapse functions, both at the particle positions. 
These quantities show small oscillations with an amplitude of about 1.5 \%, which
is larger than, for example, in \cite{tootle21}. These residual oscillations likely result
from a non-perfect translation of the FUKA initial data to our mixed
adaptive-mesh-particle methodology. This will be further refined in future work.
In run \texttt{SLy\_14\_14}, the 
density starts to increase monotonically after contact, while the lapse begins to drop. When the first particles 
are removed with lapse below $\alpha_{\rm cut}$ (causing the  floor value in the green curve in the right panel), 
the densities are greater than $10^{17}$ \gcc. Once the dense core has been devoured by the newly
formed black hole, the peak densities drop towards zero; see
the left panel. None of the other cases seems to be endangered by
imminent black hole formation, but in some cases the lapse keeps decreasing at a slow pace
(e.g. for \texttt{MPA1\_12\_18} and \texttt{SLy\_13\_13}), so 
black holes could form at later stages.

%%%%%%%%%%%%%%%%%%%%%%%%%%%%%%%%%%%%%%
%%%%%%%%%%%%%%%%%%%%%%%%%%%%%%%%%%%%%%

\section{Ejection mechanism}
\label{sec:mechanism}
\subsection{Generic case: equal masses, surviving remnant}
\begin{figure*}
  \centerline{
  \includegraphics[width=2\columnwidth]{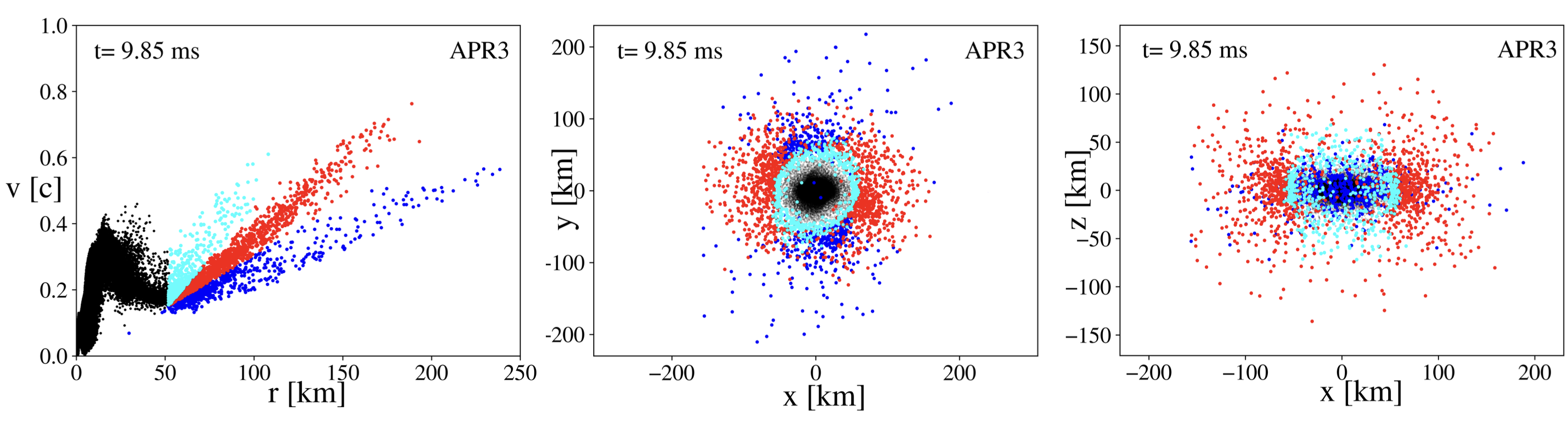} 
  }
  \caption{Identification (at $t= 9.85$ ms) of three ejection pulses for run
  \texttt{APR3\_13\_13}. We color code the three pulses visible in the left panel: the dark blue "spray ejecta" emerge
  when matter is sprayed out from the stellar interfaces at first contact. The second pulse (red) is ejected when the strongly compressed remnant bounces back 
  ("bounce ejecta"). A second, weaker pulse (cyan) emerges from a subsequent bounce. The spray component is restricted to the orbital plane, the first bounce component is rather spherical (apart from some obstruction from the earlier spray component), the second bounce component is substantially braked by the earlier ejecta and expands easiest in the polar direction. These ejecta pulses (same colouring) are followed through the hydrodynamic evolution in Fig.~\ref{fig:vel_branches_APR3_evolution} .}\label{fig:velocity_branches_APR3_2x1.3}
\end{figure*}

While mass ejection in neutron star mergers has been studied for quite some time,
we want to give it a fresh look here. When plotting the matter velocity as a function
of radius, we typically find several ejecta "branches" that are launched one after another; see
Fig.~\ref{fig:velocity_branches_APR3_2x1.3} for a typical case with 2 $\times 1.3$ \Msun and
the APR3 EOS. This case produces a central remnant that does not show any sign of collapse during the simulation
and we see three, maybe four, such velocity branches. Since, at the time of the snapshot, most of the elements in the branches are homologous, at any given time $t$, each element satisfies $r \approx v(t-t_{\rm ej})$, where $t_{\rm ej}$ is the time at which the element was ejected. Therefore, each branch corresponds to a separate ejection episode where steeper branches have later $t_{\rm ej}$. In the right two panels of Fig.~\ref{fig:velocity_branches_APR3_2x1.3}, we also show
the location of these particles (projection onto the $XY$- and $XZ$-plane) at the moments when we identify them ($t=9.85$ ms).\\
%%%%%%%%%%%%%%%%%%%%%%%%%%%%%%%%
\begin{figure*} %  figure placement: here, top, bottom, or page
   \centerline{
   \includegraphics[width=3in]{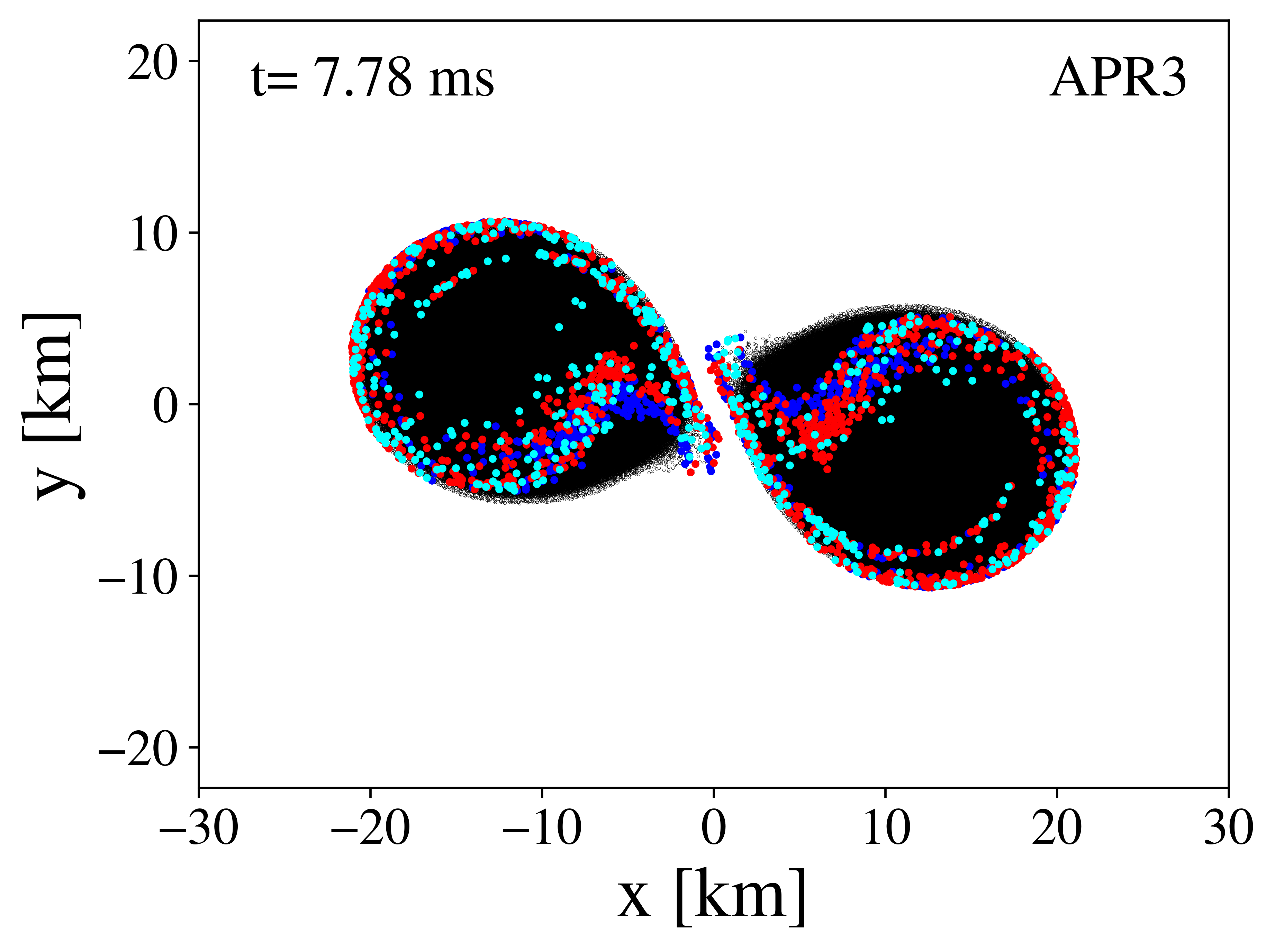} 
   \includegraphics[width=3in]{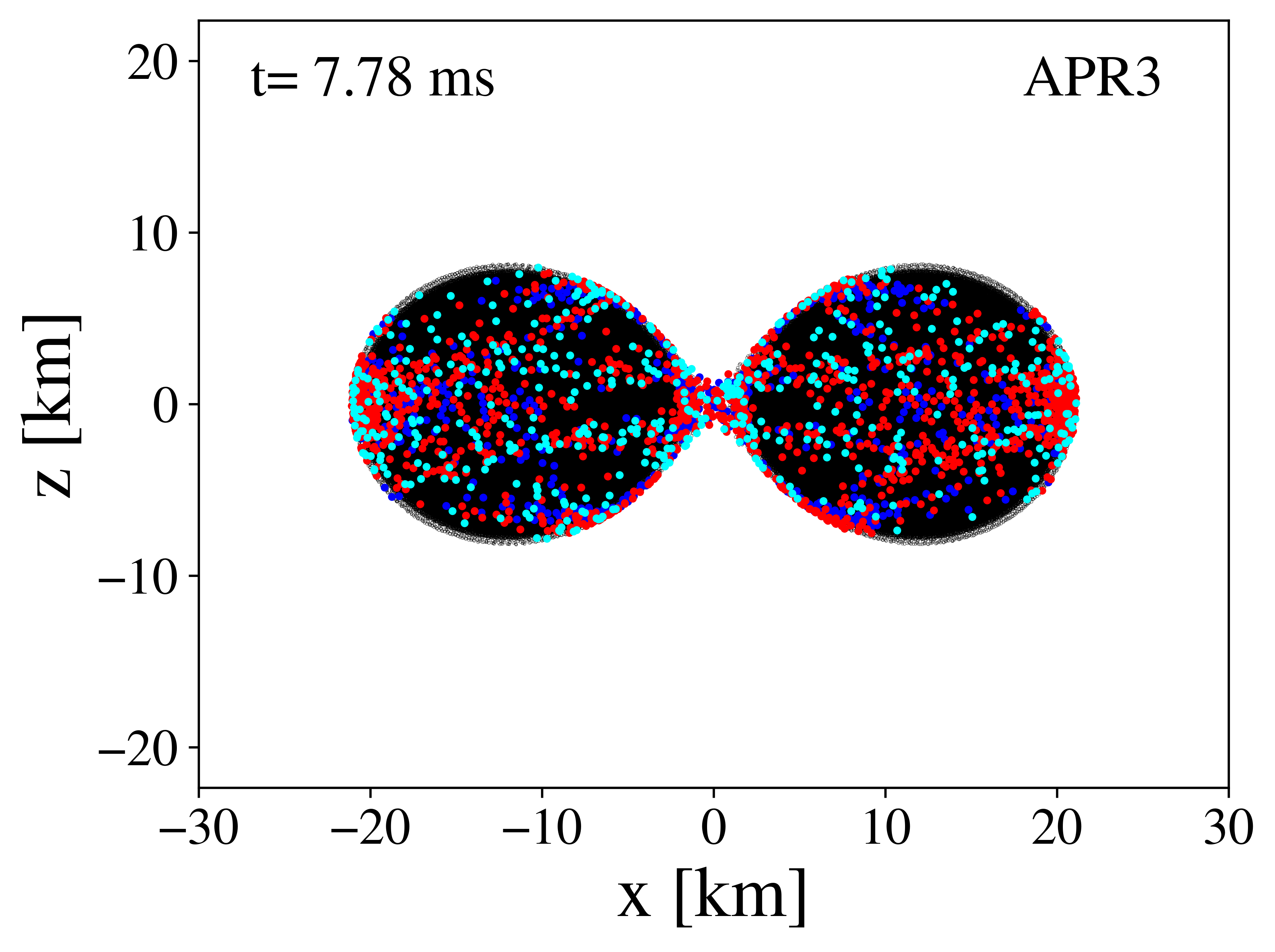} }
   \centerline{
   \includegraphics[width=3in]{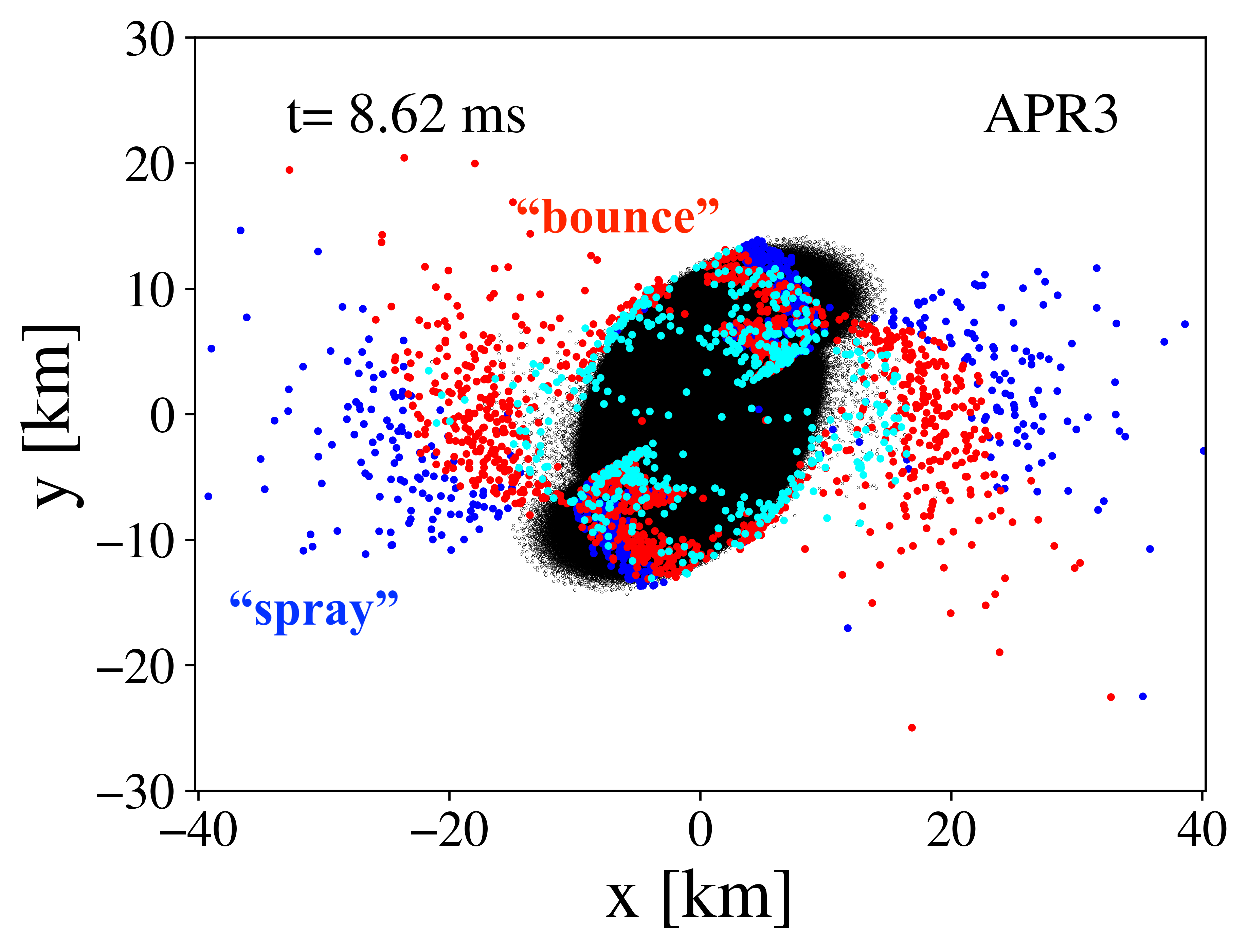} 
   \includegraphics[width=3in]{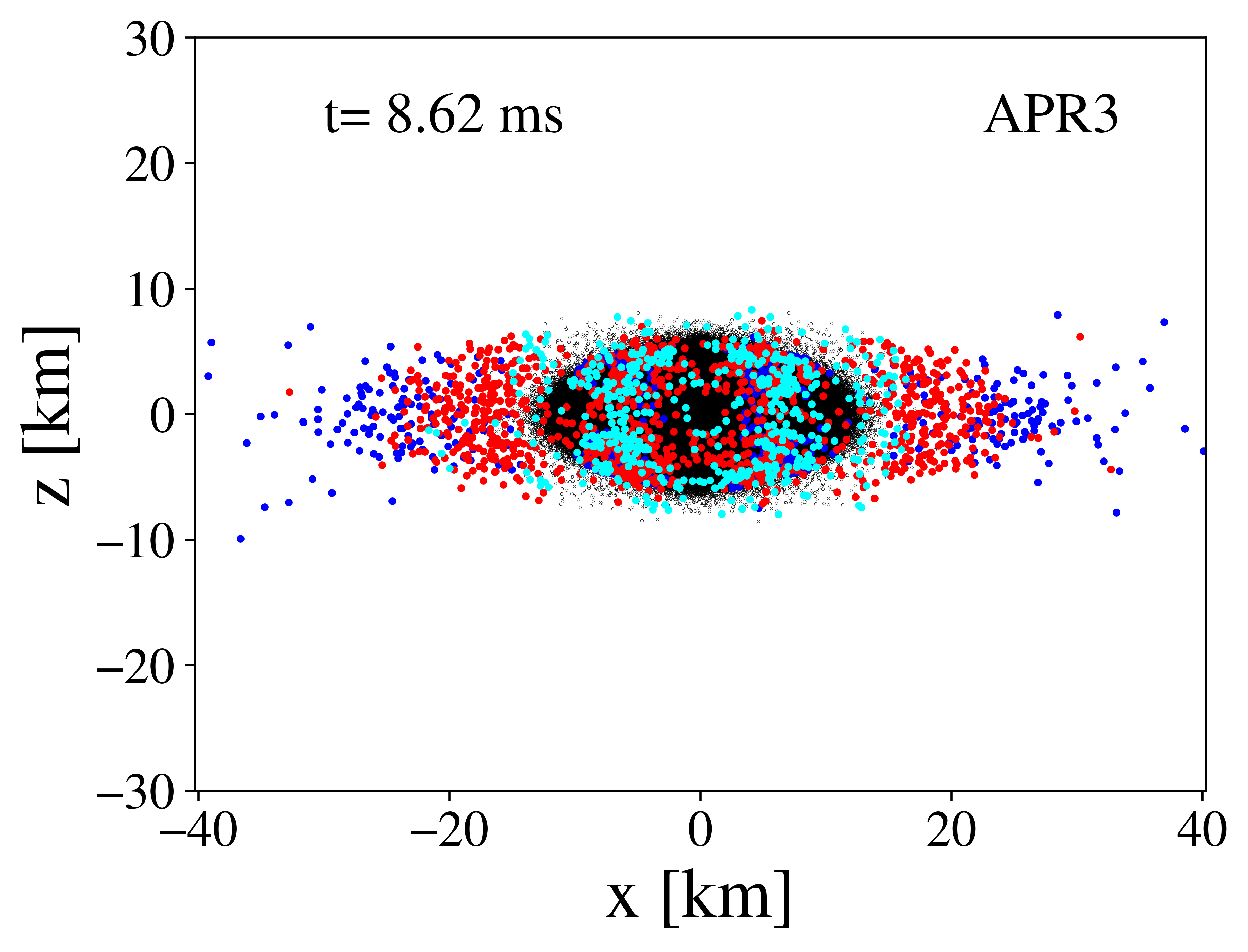} }
  \centerline{
   \includegraphics[width=3in]{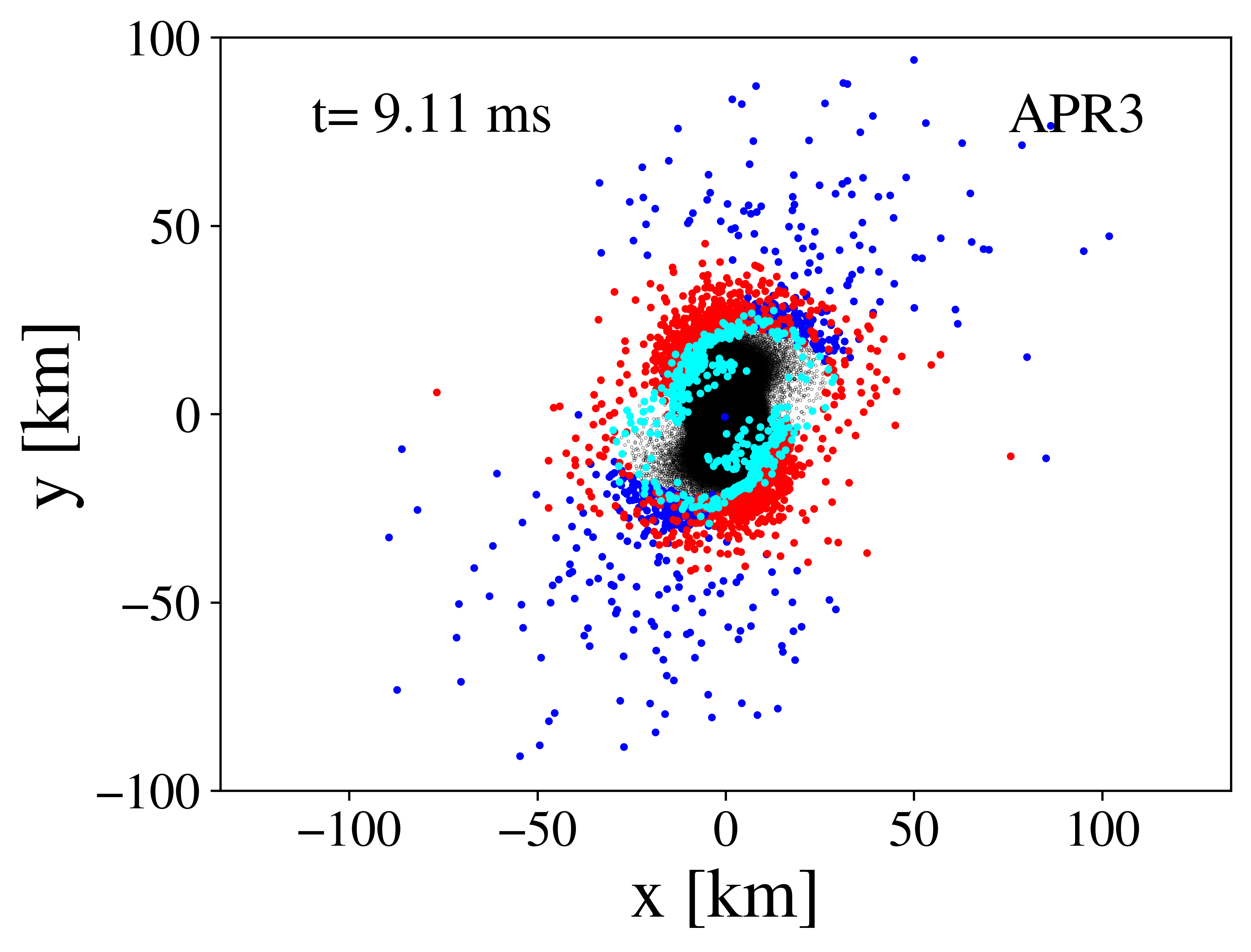} 
   \includegraphics[width=3in]{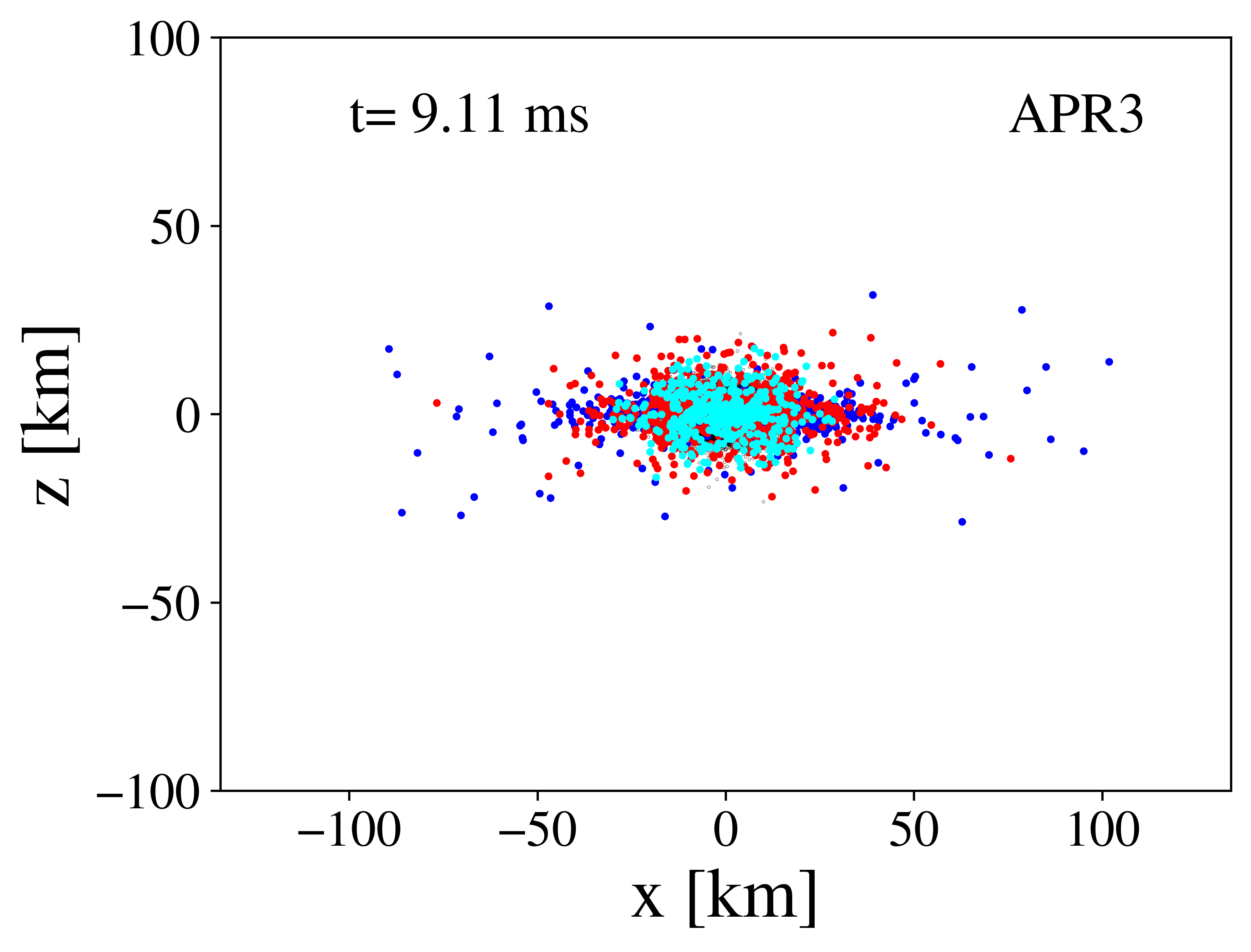} }
   \centerline{
   \includegraphics[width=3in]{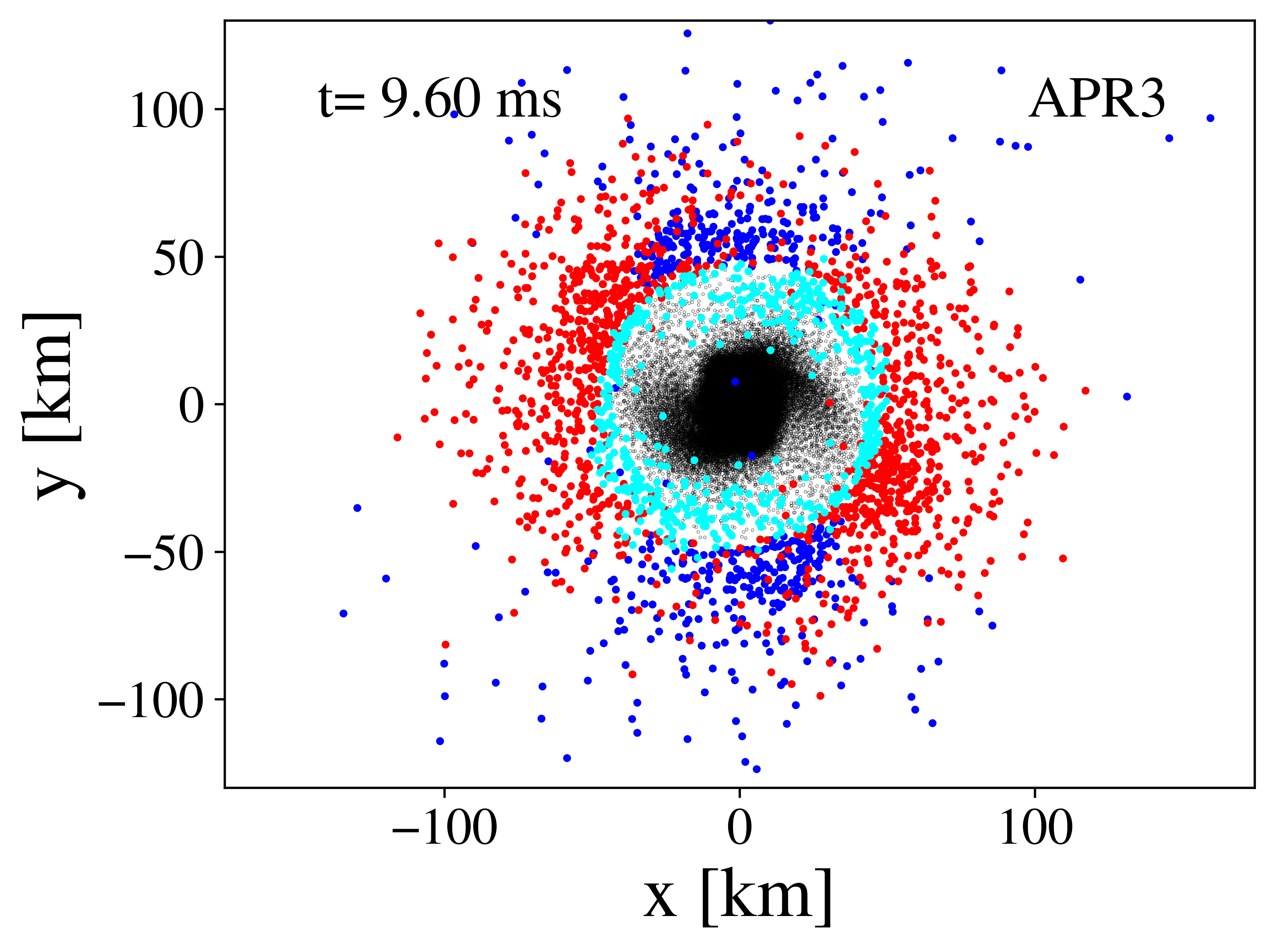} 
   \includegraphics[width=3in]{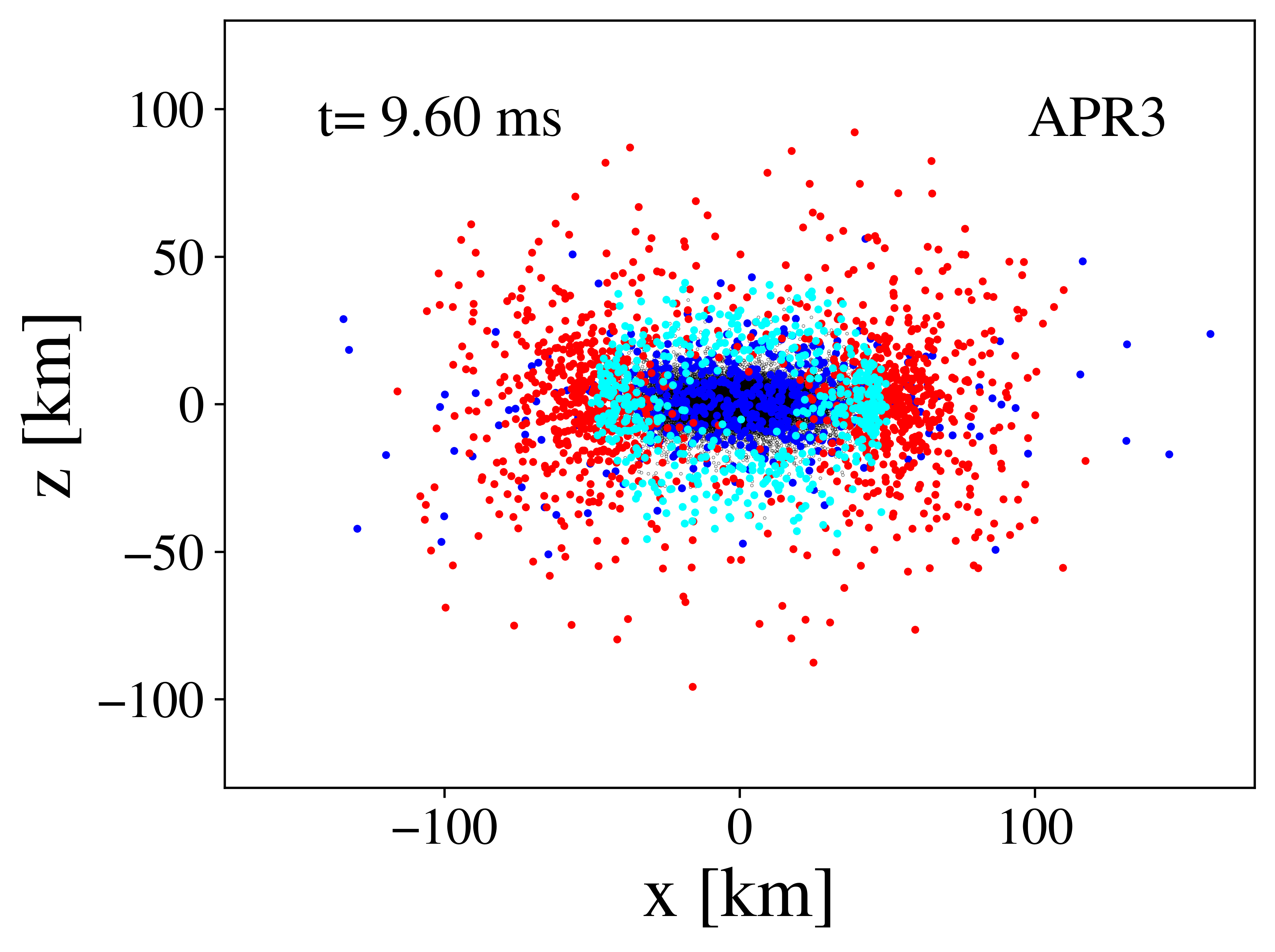} }   
   \caption{Particles of run \texttt{APR3\_13\_13} are color-coded according to the velocity waves they belong to, see Fig.~\ref{fig:velocity_branches_APR3_2x1.3}: the "spray-component" (dark blue) is ejected immediately from the interface, while
   the first (red) and  second (cyan) "bounce components" initial stay close to the remnant, but then are ejected when the remnant bounces back.}
\label{fig:vel_branches_APR3_evolution}
\end{figure*}
%%%%%%%%%%%%%%%%%%%%%%%%
%
%%%%%%%%%%%%%%%%%%%%%%%%
\begin{figure*} %  figure placement: here, top, bottom, or page
   \centerline{
   \includegraphics[width=3.5in]{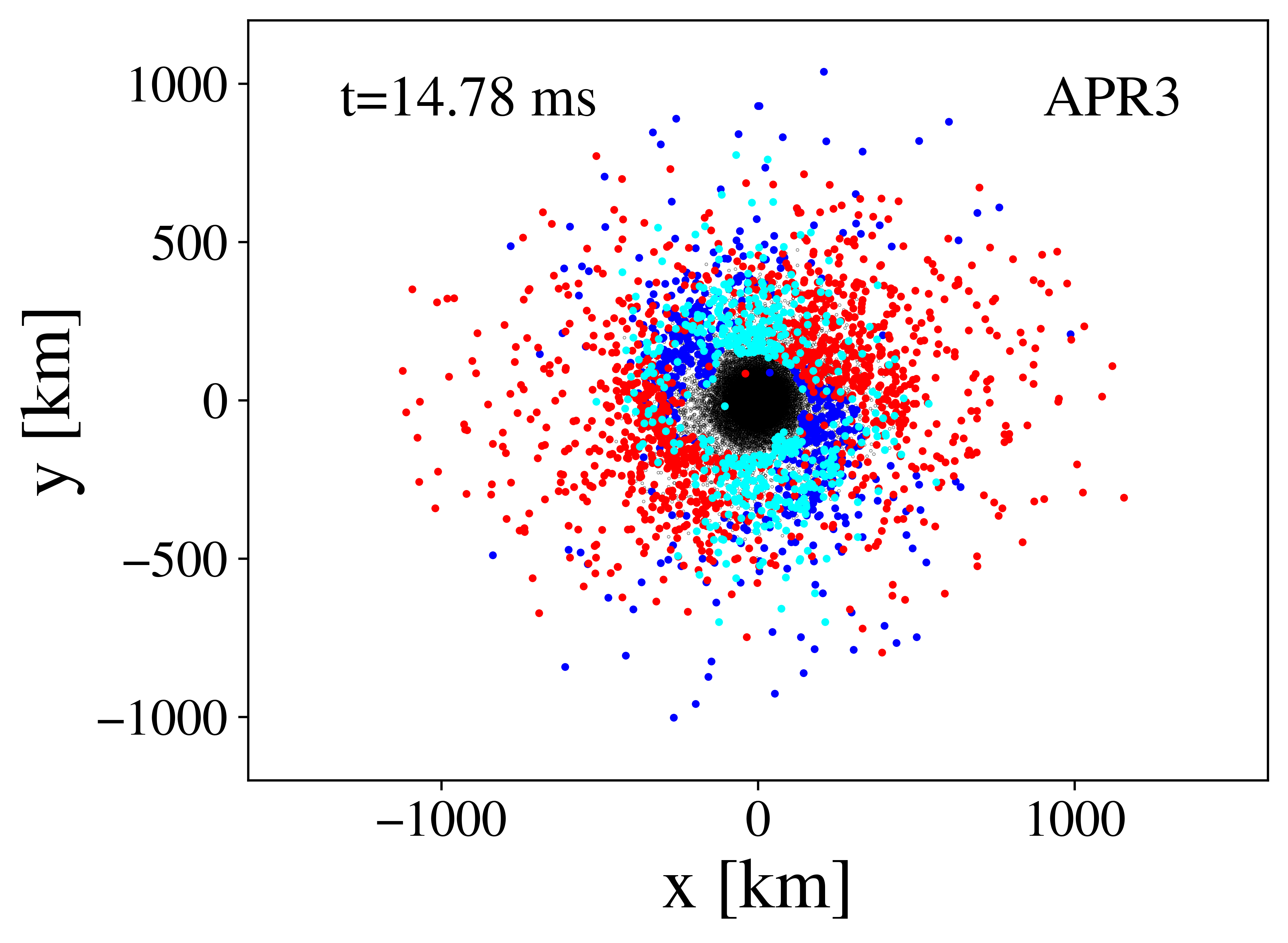} 
   \includegraphics[width=3.5in]{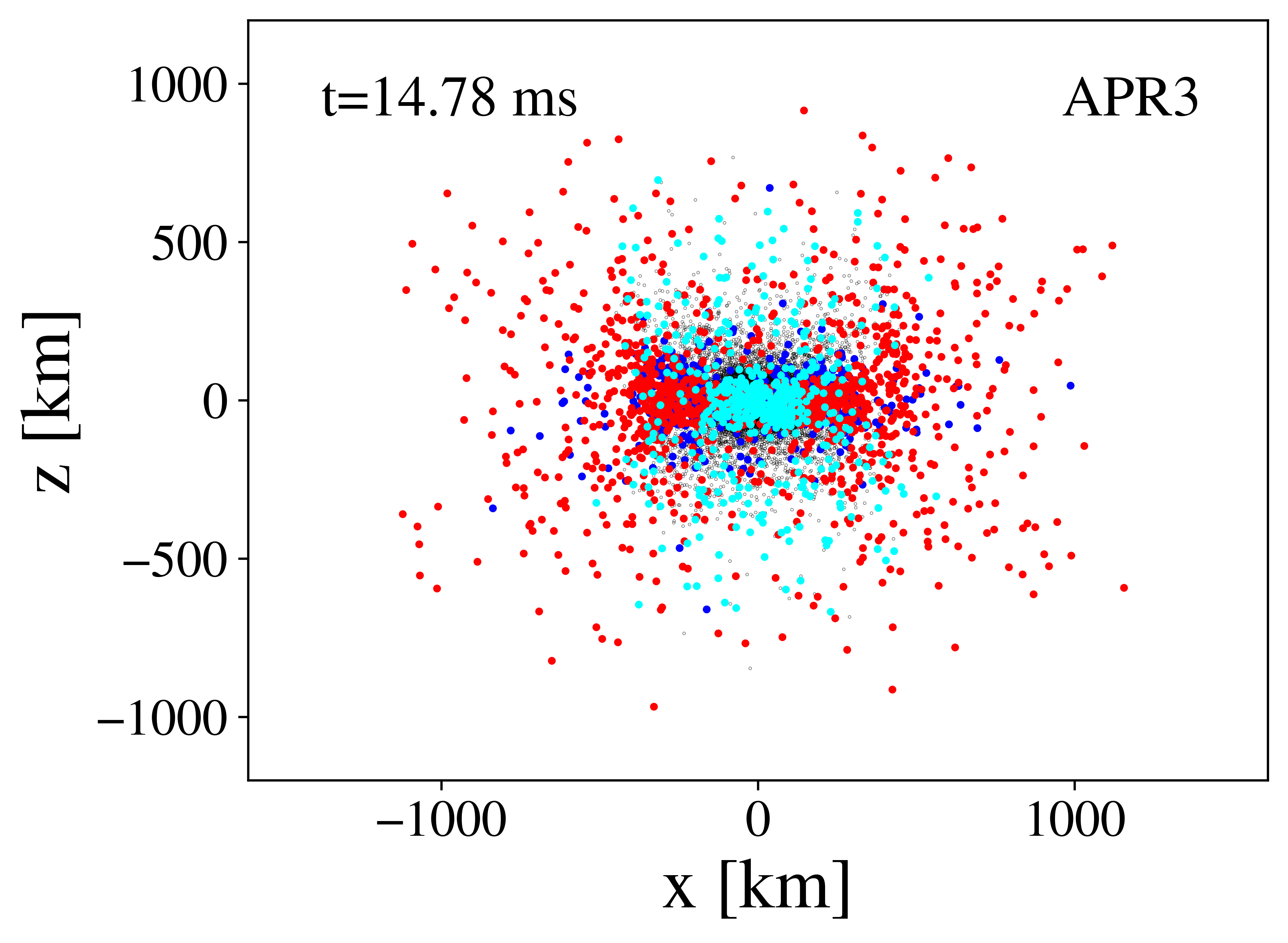} }
   
    \caption{The spray component (dark blue) is launched in predominantly equatorial direction, the two bounce components
    are launched predominantly spherically, but their expansion is hindered by the previously launched wave(s). The colours here indicate the different ejection components as in Fig.~\ref{fig:vel_branches_APR3_evolution}.}
   \label{fig:APR3_waves_2}
\end{figure*}
%%%%%%%%%%%%%%%%%%%%%%%%
%
%%%%%%%%%%%%%%%%%%%%%%%%
\begin{figure*} %  figure placement: here, top, bottom, or page
   \centerline{
   \includegraphics[width=3in]{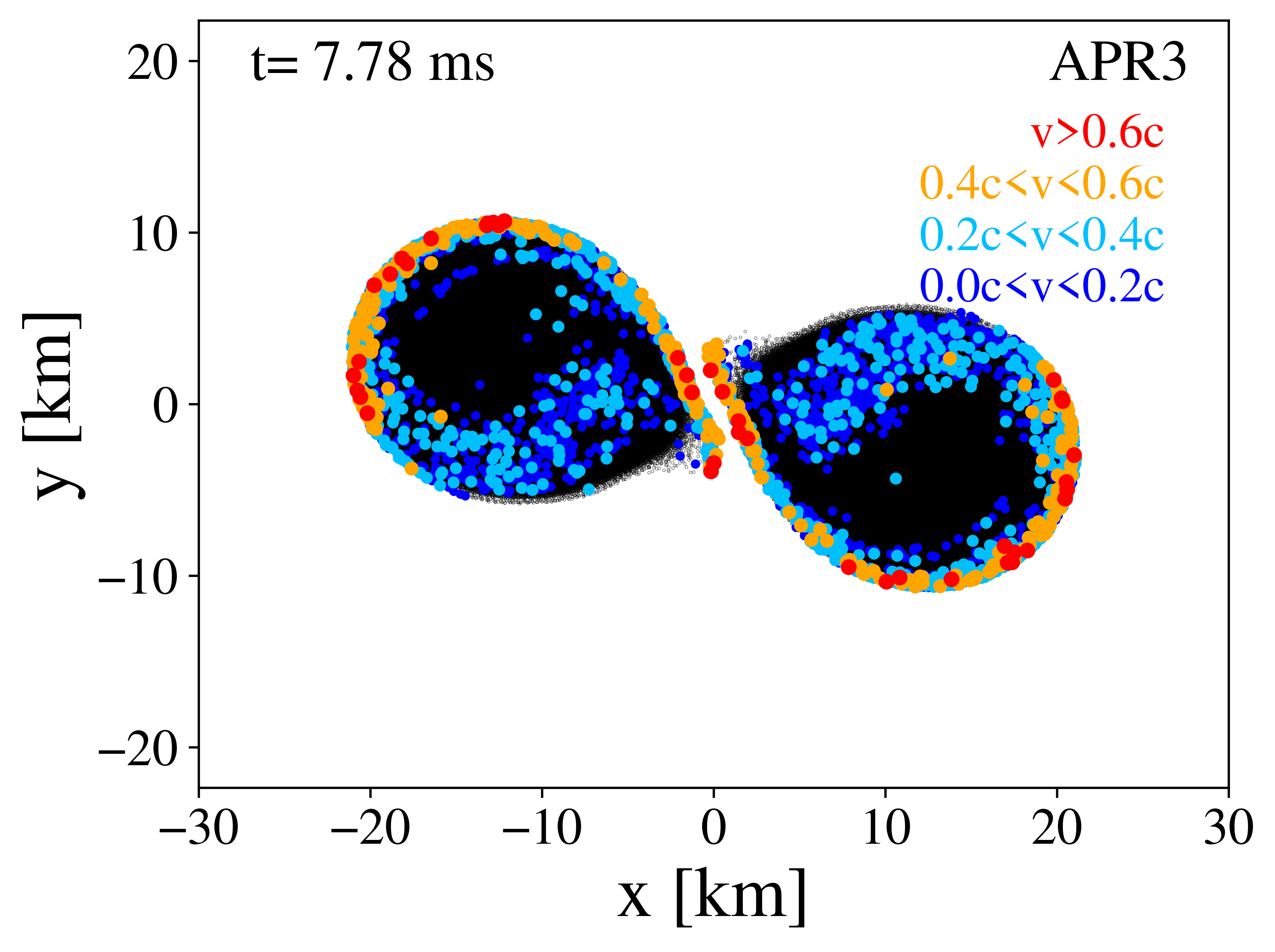} 
   \includegraphics[width=3in]{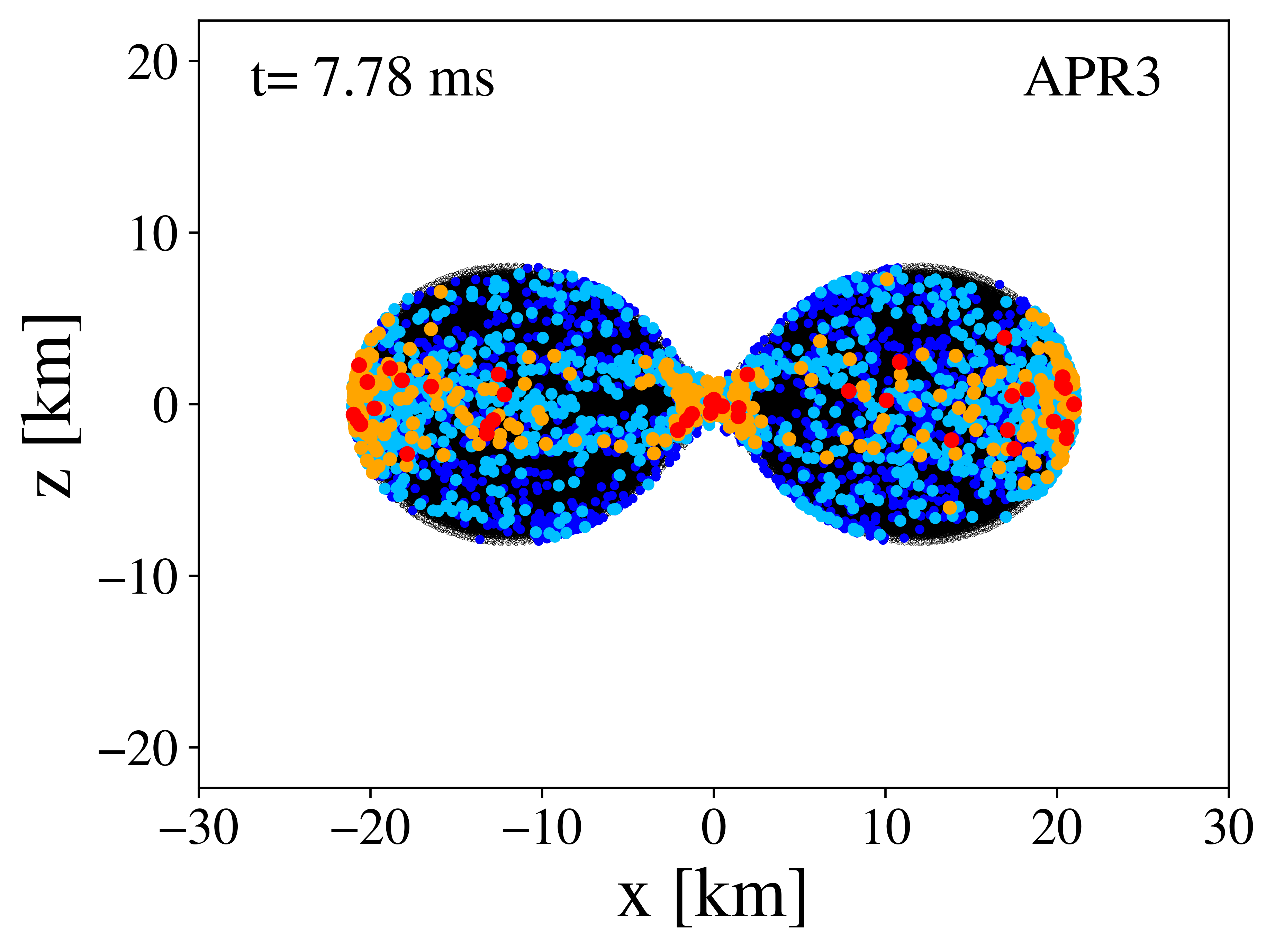} }
   \centerline{
   \includegraphics[width=3in]{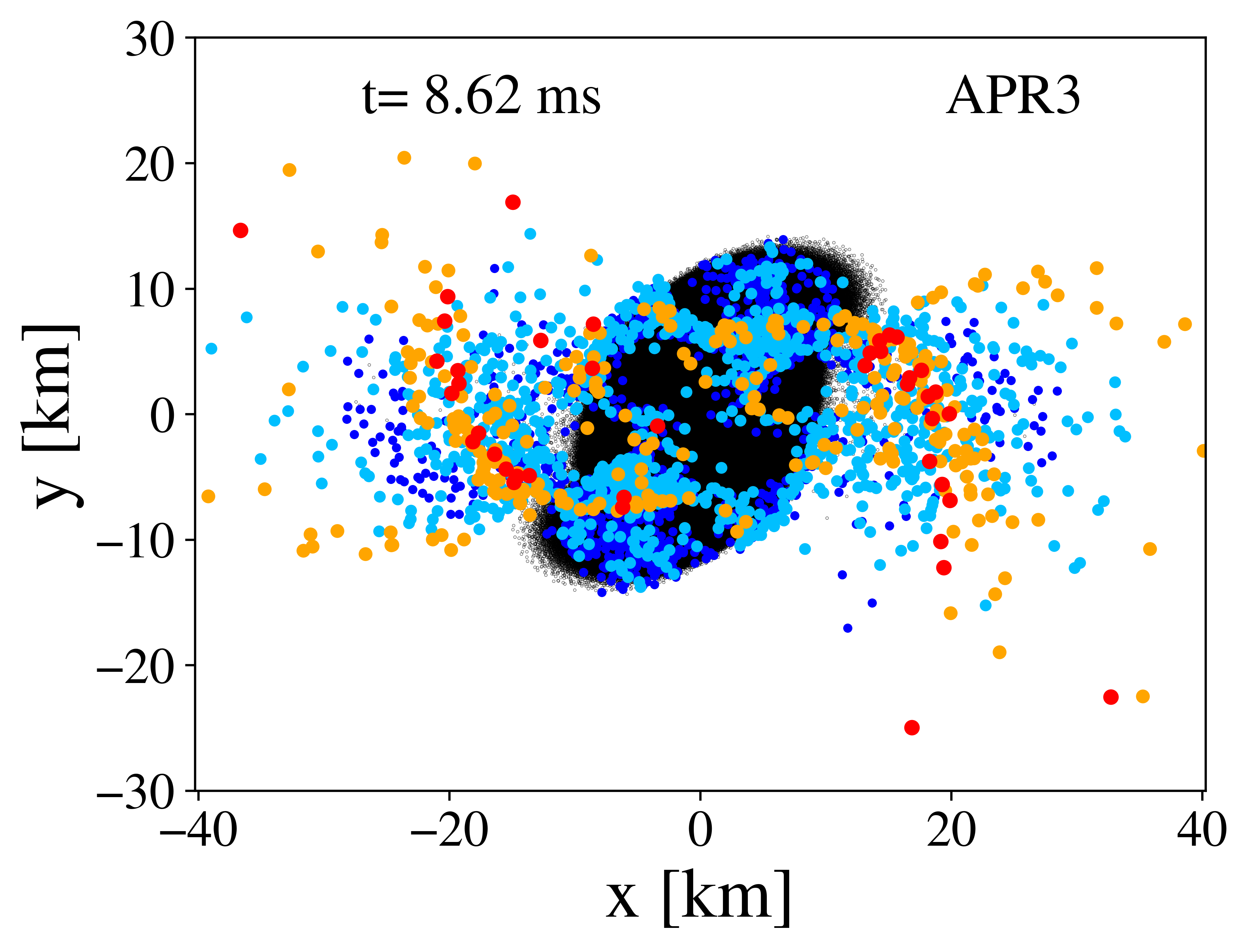} 
   \includegraphics[width=3in]{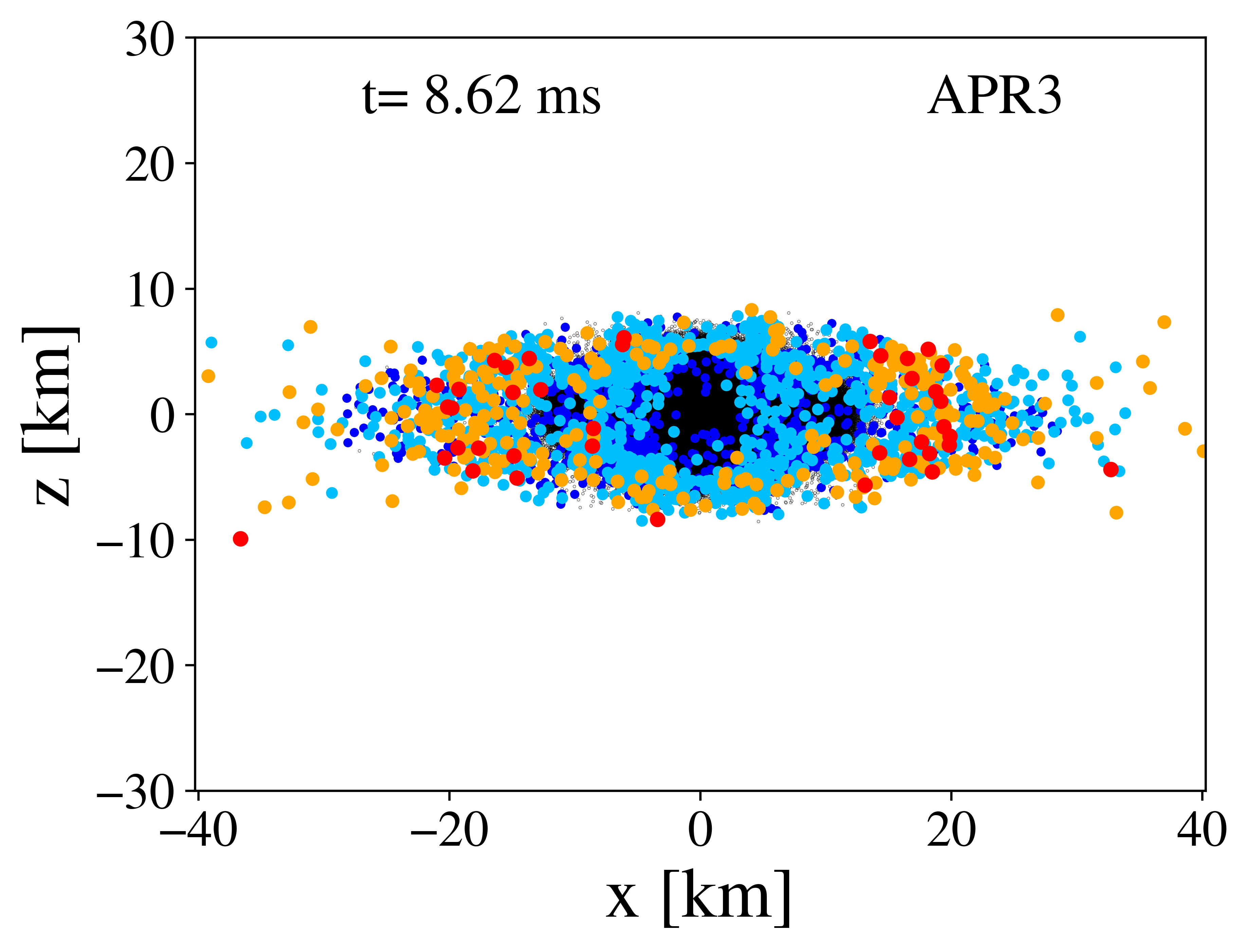} }
  \centerline{
   \includegraphics[width=3in]{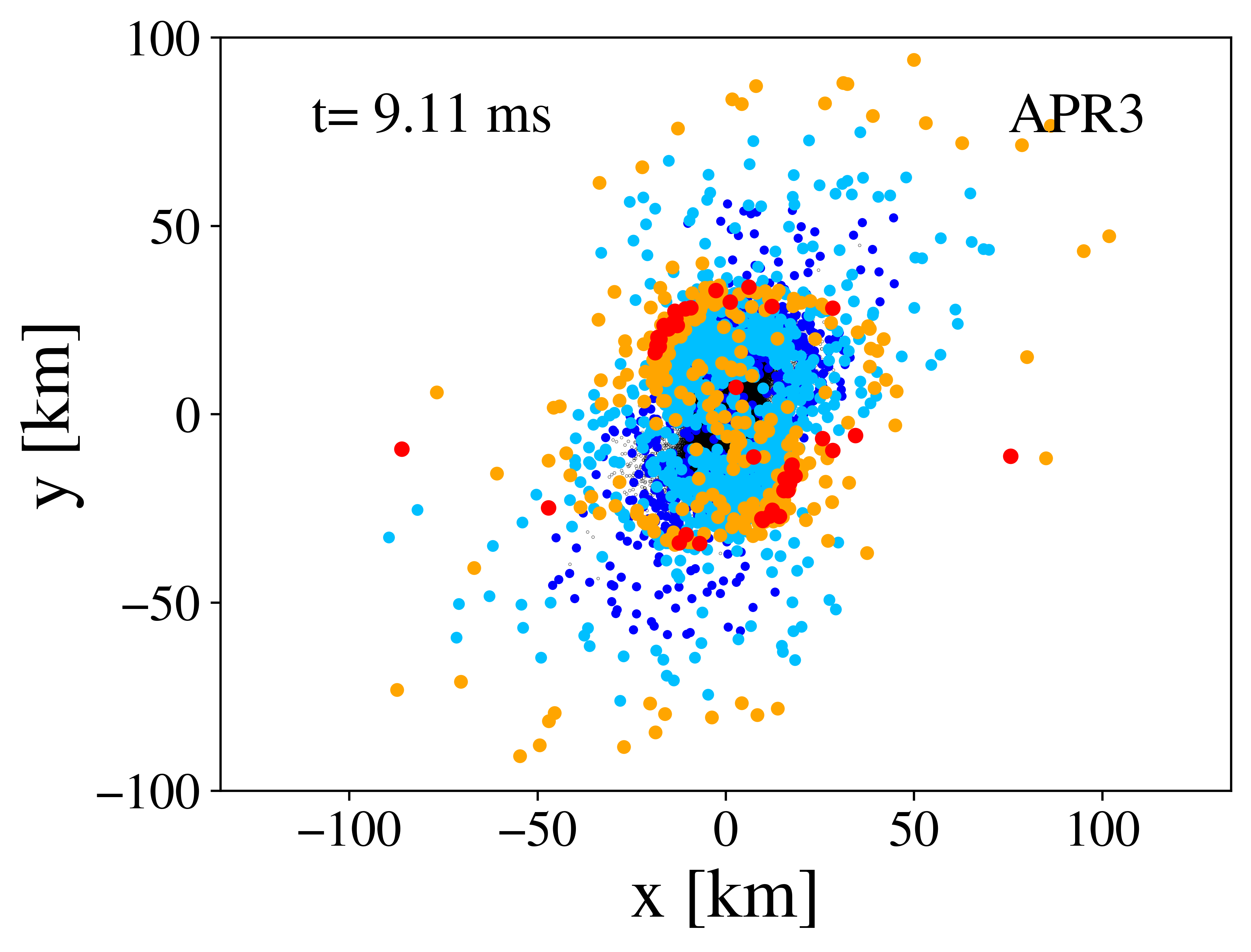} 
   \includegraphics[width=3in]{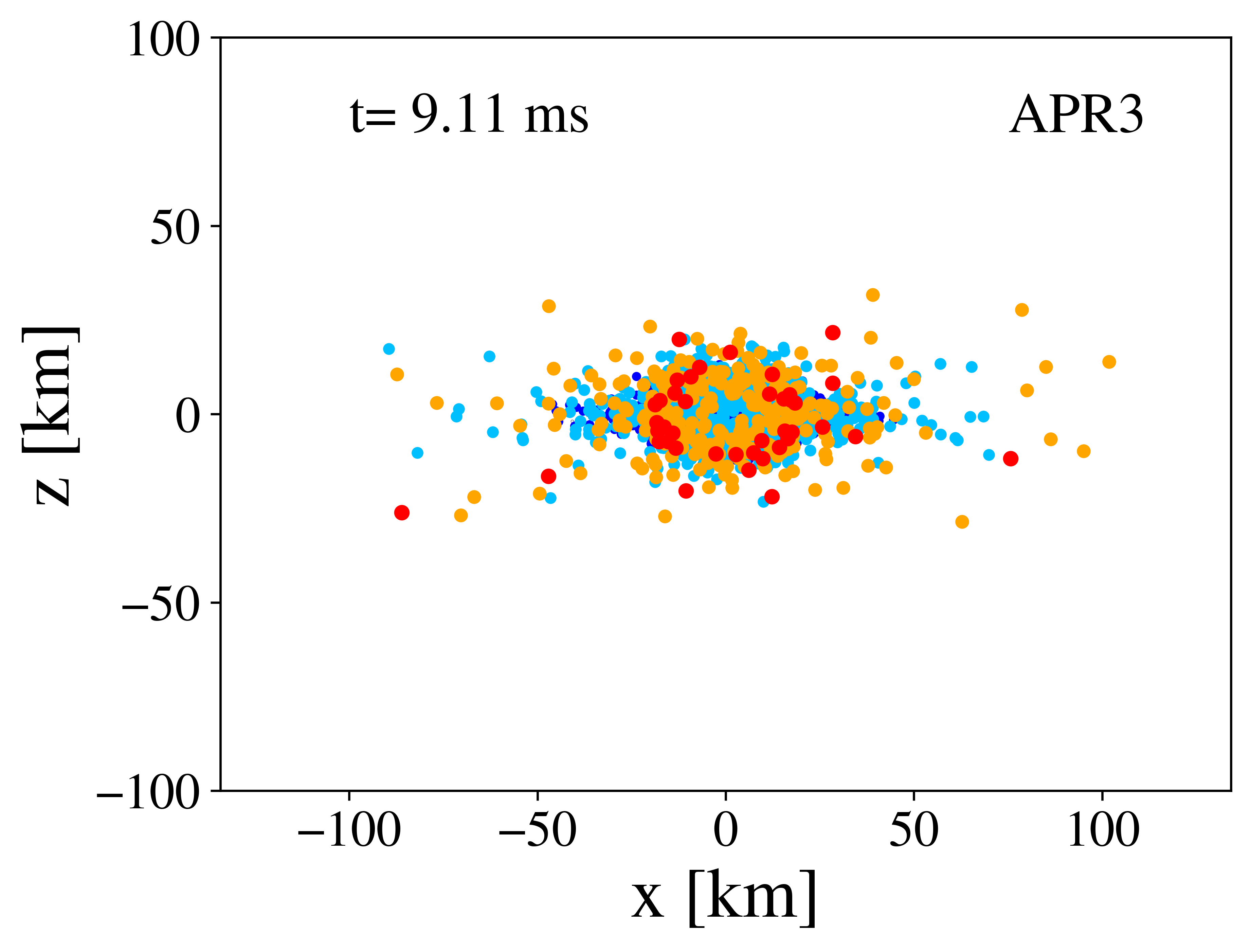} }
   \centerline{
   \includegraphics[width=3in]{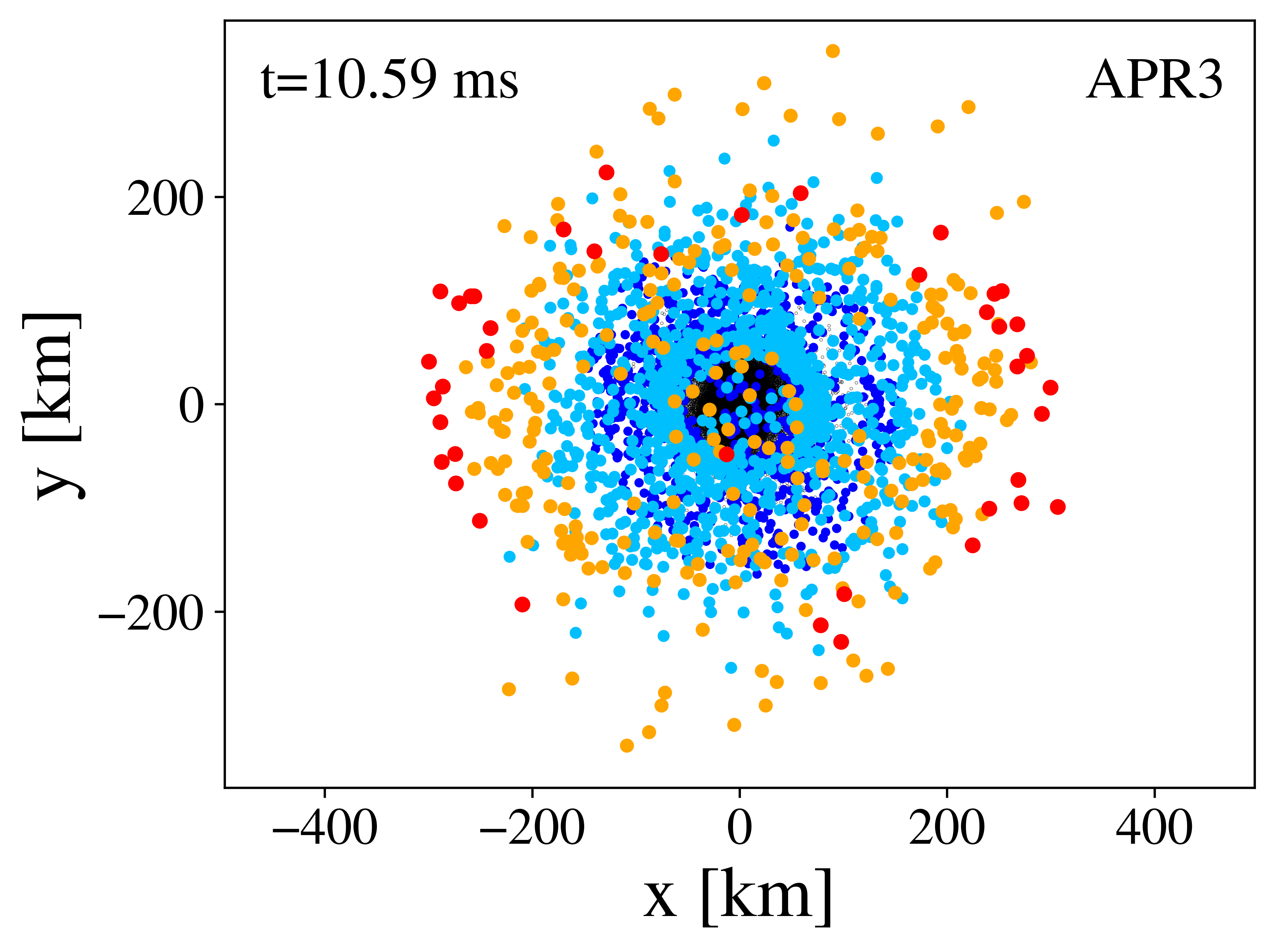} 
   \includegraphics[width=3in]{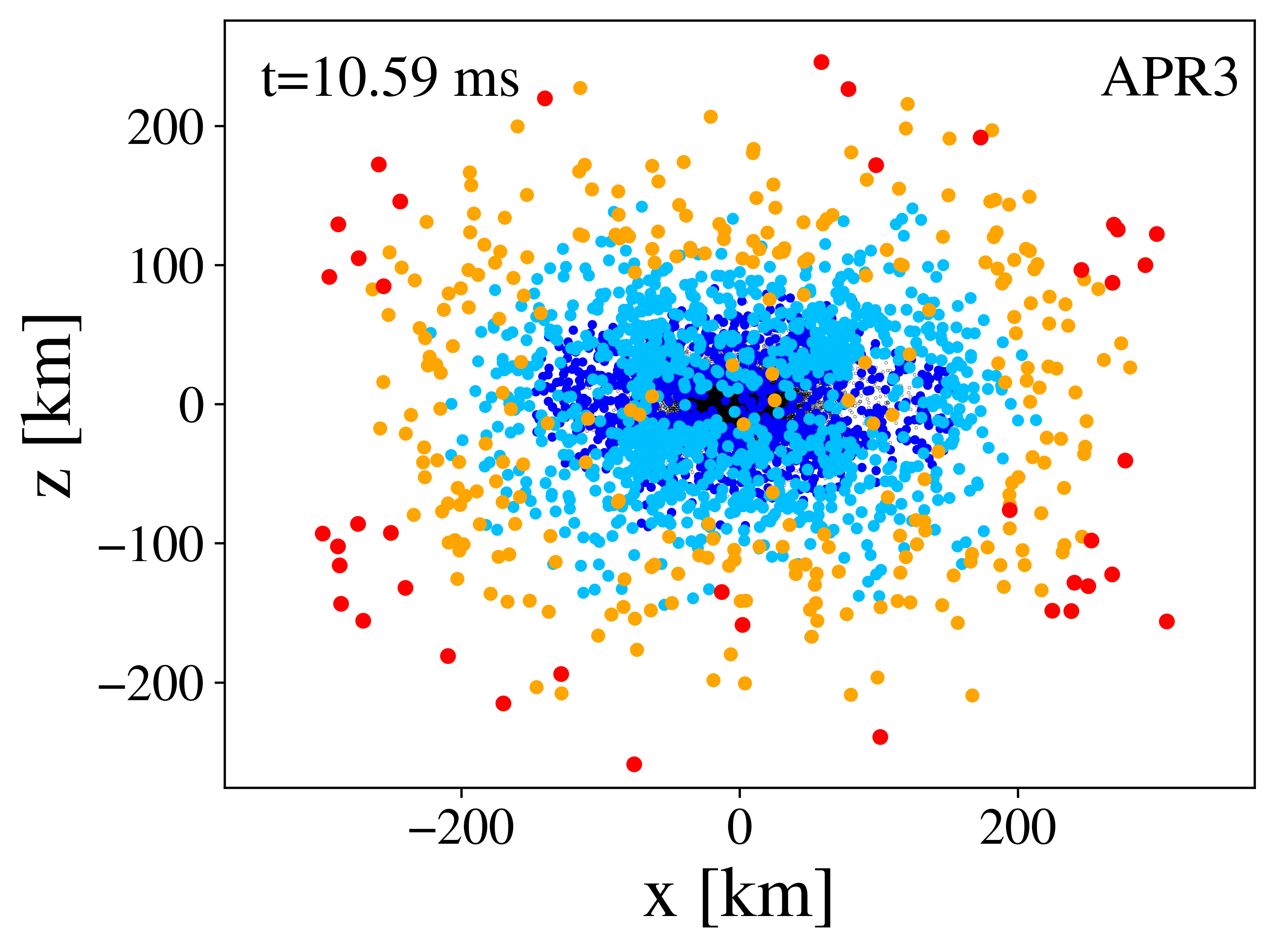} }   
   \caption{Unbound particles are color-coded according to their velocities at infinity: ejecta slower that $0.2c$ (at infinity) are shown in dark blue,particles between 0.2 and 0.4$c$ in cyan, between 0.4 and 0.6$c$ in orange and the fastest ones ($>0.6c$) are shown in red.}
\label{fig:vel_bins_APR3_evolution}
\end{figure*}
%%%%%%%%%%%%%%%%%%%%%%%%
In Fig.~\ref{fig:vel_branches_APR3_evolution} we follow the  particles in these branches (same color-coding) throughout
their evolution.
All of the ejected particles are initially rather evenly spread across the initial neutron stars, see panel 2 in Fig.~\ref{fig:vel_branches_APR3_evolution}. At contact (panel three and four), particles are "sprayed out" from the interface between the two stars, predominantly into the orbital plane. Their leading part (colored in dark blue) immediately moves away from the remnant, while branch two (red) and three (cyan) still stay in the vicinity of the remnant. The initial contact between the two stars leads to a deep compression of the central remnant, see the left panel in Fig.~\ref{fig:rho_lapse}, and a subsequent strong "bounce back" which
expels the particles marked in red, while the next bounce ejects the particles marked in cyan. In the following, we refer to the first wave (dark blue) as "spray component", and the subsequent ones (red and light blue) as "bounce components". Typically, the first bounce produces the highest velocity ejecta and subsequent bounces become increasingly weaker.\\
The spray component (dark blue) is mostly restricted to the orbital plane, while the first bounce component (red) is rather spherical (apart from some obstruction along the orbital plane from the earlier spray component) and the second bounce component is substantially braked by the earlier ejecta and expands easiest along the polar directions, see Fig.~\ref{fig:APR3_waves_2} for the same simulation, same colour coding, at a later stage (14.78 ms).
Since the first bounce ejecta catch up with the spray ejecta
in the orbital plane, they are braked along the orbital plane, but not perpendicular to it,
see panel two in Fig.~\ref{fig:APR3_waves_2}, while the
second branch (cyan) is additionally braked by the slower 
ejecta portion of the second branch and therefore predominantly
expands along the rotation axis.
This finding is also sketched in Fig.~\ref{fig:vfast_APR3_v3} of the summary section.\\
In Fig.~\ref{fig:vel_bins_APR3_evolution} we show the same simulation (\texttt{APR3\_13\_13}), but this time with a different colour coding.
Rather than colouring according to their ejection mechanism, we now colour-code the particles according to their asymptotic velocity: particles up to 0.2c are shown in dark blue,
particles between 0.2 and 0.4c are shown in cyan, between 0.4 and 0.6 in orange and the 
fastest particles ($>0.6c$) are shown in red.  Interestingly, while the slower particles ($<0.4c$)
are initially spread  relatively uniformly over the stellar surfaces, the faster ejecta ($>0.4c$)
come from equatorial belts.

\subsection{Unequal mass case \texttt{MPA1\_12\_18}}
To get a first idea of the impact of the mass ratio, we look
at run \texttt{MPA1\_12\_18} with the same EOS, but very different stellar masses (1.2 and 1.8 \msun). Here we again find 3-4 branches 
of ejecta and peak velocities up to
0.8c. As for all of the runs presented here, we expect that higher 
resolution will lead to small amounts of ejecta reaching even larger velocities. For this case we find a smaller degree of sphericity in the ejecta, both for the bulk and the fastest parts.
Overall, the ejecta distribution is roughly lenticular. 

\subsection{EOS-dependence}
\begin{figure*} 
   \hspace*{-2cm}\includegraphics[width=2.6\columnwidth]{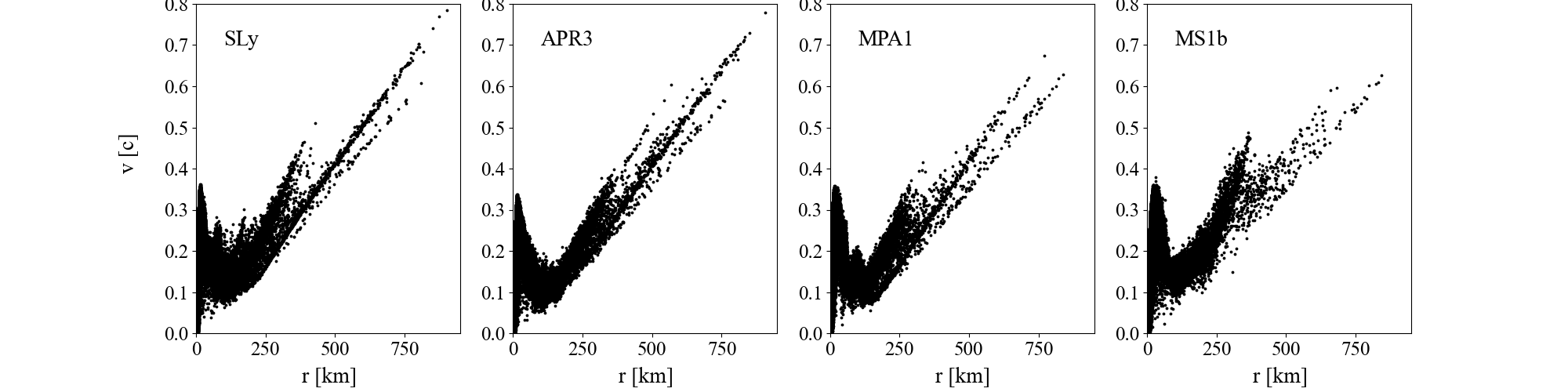}
    \caption{EOS-dependence of the ejecta velocities (in units of $c$; each time $2 \times 1.3$ \msun). Shown are each time the particle velocities at 5 ms after merger for
    the SLy (softest), APR3, MPA1 and the MS1b (stiffest) EOS.}
   \label{fig:EOS_dependence}
\end{figure*}

To illustrate the EOS-dependence, we plot the particle
velocities at $\approx$ 5 ms after the merger for the $2 \times 1.3$
\Msun systems for the different equations of state in Fig.~\ref{fig:EOS_dependence}.
%%%%%%%%%%%%%%%%%%%%%%%%%%%%%%%
%%%%%%%%%%%%%%%%%%%%%%%%%%%%%
The velocities are shown at t= 13.1 ms (SLy), t= 13.3 ms (APR3), t= 11.9 ms (MPA1) and t= 12.0 ms (MS1b). Not completely unexpected, one sees the tendency, that softer EOSs produce larger peak velocities. Also for the two softest EOS (SLy and APR3) the first "bounce"  branch (=second branch in total) of the ejecta produces the largest velocities, 
while its peak velocities become comparable to the "spray" component
for the stiffer EOSs (MPA1 and MS1b).\\
\subsection{Impact of total binary mass}
To get a qualitative idea of the impact of the total mass
on the peak velocities, we compare the velocities of the $2 \times 1.3$ \Msun systems with those of the $2 \times 1.4$ \Msun cases
for our four EOSs. Since these are just a few cases this comparison should be taken with a grain of salt. For the SLy EOS, we find a difference, in the peak velocities between the 2 $\times$ 1.4 and the 2 $\times$ 1.3 \Msun case, of $\Delta v_{1.4\_1.3}\approx 0.04 c$, for MPA1 we find $\Delta v{1.4\_1.3}\approx 0.16 c$, no significant difference for APR3
and $\Delta v_{1.4\_1.3}\approx 0.02 c$ for MS1b. So there is a tendency for
faster peak ejecta with higher masses, but at least for the small mass
range that we have explored here, the differences are moderate
apart from the MPA1 case.
%%%%%%%%%%%%%%%%%%%%%%%%%%%%%%%
\begin{figure} 
   \includegraphics[width=1.1\columnwidth]
   {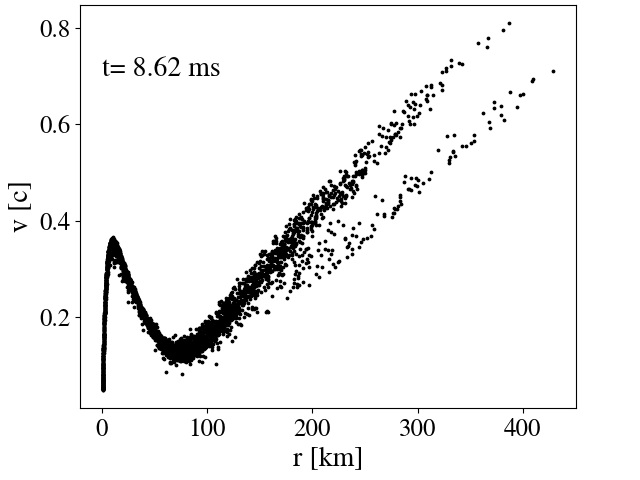}   %{velocity_branches_prompt_collapse.pdf}
    \caption{Even the case which results in a prompt collapse to a black hole, run \texttt{SLy\_14\_14}, produces fast ejecta. As in the other cases, the first branch is due to the "spray component" while the second comes from
    a single "bounce". Velocities are shown in units of $c$.}
   \label{fig:vel_p_collapse}
\end{figure}
%%%%%%%%%%%%%%%%%%%%%%%%%%%%%
\subsection{Case of prompt collapse}
Interestingly, we also find fast ejecta in our prompt collapse case, run 
\texttt{SLy\_14\_14}, where the central
density increases monotonically  and  where a black hole forms on a free-fall time, see Figs.~\ref{fig:prompt_collapse} and \ref{fig:rho_lapse}.
Here we only see two branches, see Fig.~\ref{fig:vel_p_collapse}, a first "spray" branch
(up to $0.7c$) similar to the ones described before and --while the innermost parts continuously contract towards collapse-- the outer layer manage to bounce back 
and eject a second "bounce" branch that reaches up to $0.8c$. Thus, even this prompt collapse case ejects
high-velocity material. While this is
the only example of a promptly collapsing central remnant among the simulations in Tab.~\ref{tab:runs},
we have additionally inspected other simulations where a black hole forms on a dynamical time scale, and find also there high-velocity ejecta. So fast ejecta
may be a wide-spread feature even for  prompt collapse cases. This issue 
deserves  further systematic study.\\

\subsection{Comparison with other work}
Our findings of a leading spray component followed by several bounce
components is broadly consistent with the findings of \cite{radice18a} who
find, similar to us, a smooth distribution of masses out to high velocities
and that the bulk of the high-velocity ejecta is launched by the re-bouncing
central object. They also see a substantial spread in the masses of high-velocity ejecta,
but overall smaller amounts, of order a few $10^{-6}$ \Msun with $v_\infty>0.6c$, 
while we find roughly a few $10^{-5}$ \msun, see Tab.~\ref{tab:ejecta_props}. One potential reason for this 
difference could be their use of a rather large "vacuum" background density of
$6 \times 10^{4}$ g cm$^{-3}$. \cite{combi23} find in two of their three cases
two ejection pulses which could potentially be what we refer to as spray and (first)
bounce ejecta. Their simulations have a lower "vacuum" density than \cite{radice18a} ($600$ g cm$^{-3}$), but they still find lower amounts of fast ejecta ($\sim ~5 \times 10^{-6}$ \Msun with $v_\infty >0.6c$).
The likely best resolved study of fast ejecta to date is due to \cite{dean21}.
Based on their local simulations of the shear interface in Newtonian 
self-gravity, they find that the results are converged to within 
10\% if they apply a (very high) grid resolution length of $\sim20$ m. Based on this result,
they state "This suggests that fast ejecta quantities found in existing grid-
based merger simulations are unlikely to increase to the level needed to match 
particle-based results upon further resolution increases". This
is, however, not an apple-to-apple comparison because they had restricted themselves to axisymmetry and to 
Newtonian self-gravity. While this is a useful study that sheds
light on the needed resolution, their two strong approximations 
do not allow for a quantitative study of the re-bounce phenomenon which
is a) intrinsically three-dimensional and b) much stronger in GR than in 
Newtonian gravity.\\
We are not aware of any report in the literature that fast ejecta have been seen in the case of a prompt collapse.

\section{Ejecta properties and observational implications}
\label{sec:properties}
\begin{figure*}
\centerline{\includegraphics[width=2\columnwidth]{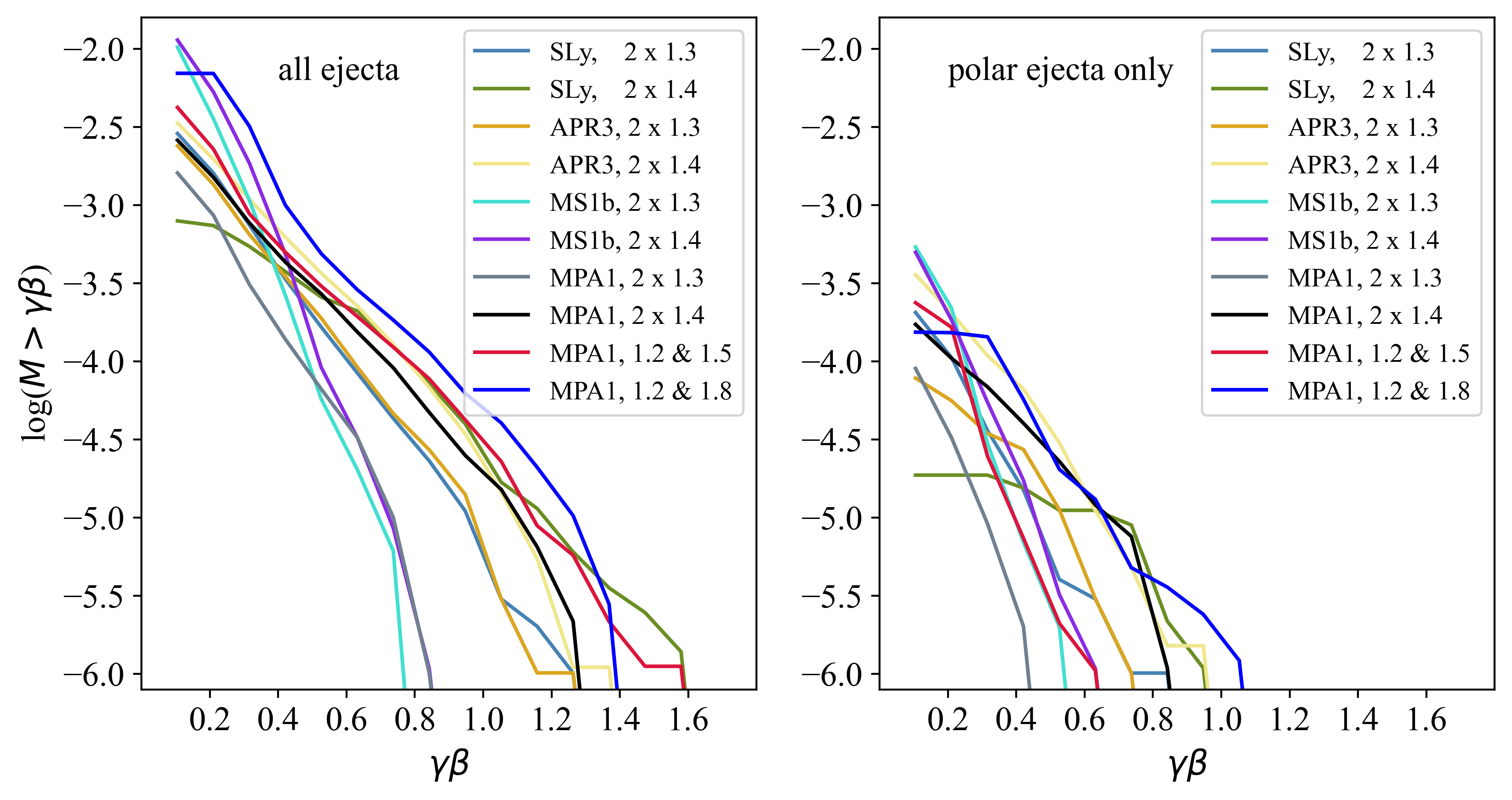}}
  \caption{Logarithm of the mass (in solar units) of the ejecta that has a velocity $> \gamma \beta$. The left panel shows the results for {\em all} ejecta, the right panel shows
  the results for the {\em polar} region (defined as being within 30$^\circ$
  of the binary rotation axis).}  
  \label{fig:M_gamma_beta}
\end{figure*}

After having discussed the ejection {\em mechanism} 
in the previous section, we now summarize the ejecta {\em properties}. Some bulk properties are summarized in Tab.~\ref{tab:ejecta_props}. 
To identify unbound matter we apply three criteria that need to be fulfilled: i) $-\mathcal{E} U_0 > 0$,  where $\mathcal{E}= 1 + u + P/\rho$ is the specific enthalpy and
$U_0$ is the time component of the four-velocity, ii) the radial velocity must be positive, $v_{\rm rad} > 0$, and a particle $a$'s radius needs to be $r_a > 100$ code units ($\approx 150$ km).
Typically, the amount of dynamic ejecta is of the order $10^{-3}$ \Msun and all cases show
fast velocities ($>0.4c$), even in
the case of \texttt{SLy\_14\_14}, where a black hole forms promptly. While this case has the smallest amount of ejecta mass, the ejecta show the largest average 
velocities ($>0.4 c$), since the slow  parts are devoured by the black hole, but the high velocity parts are very similar to the other cases. For the SLy, APR3 and MPA1 EOS, the tendency is that
softer EOSs produce more ejecta, and for the non-collapsing   cases
the heavier systems eject more matter. As expected, unequal 
mass systems eject more material and at a slightly larger
average velocity. The (probably unrealistically) stiff 
MS1b EOS ejects more than a percent of a solar mass, even
for equal mass binaries.  We find that $\sim 30$ \% of the fast ejecta ($>0.4c$) are due to the spray mechanism while the remaining $
\sim70$ \% are due to the bounce mechanism.\\
\begin{table*}
	\centering
	\caption{Properties of the dynamic ejecta for all simulations. The quantities $m_{\rm ej,X}$/$E_{\rm kin,X}$ refer to the ejecta mass/kinetic energy of matter with $v_\infty>X c$.}
	\label{tab:ejecta_props}
	\begin{tabular}{clccccccc} % four columns, alignment for each
		\hline
		run & $m_{\rm ej}$ & $\langle v\rangle$ & $E_{\rm kin,all}$ & $E_{\rm kin,0.4}$ & $m_{\rm ej,0.4}$ &  $m_{\rm ej,0.5}$ &  $m_{\rm ej,0.6}$\\
		         & [$10^{-3}$ M$_\odot$] &    [c]  & [$10^{50}$ erg]  &[ $10^{49}$ erg]      & [$10^{-4}$ M$_\odot$] & [$10^{-4}$ M$_\odot$] & [$10^{-5}$ M$_\odot$] \\
		         
		         		 \hline
\texttt{SLy\_13\_13} & 3.7 & 0.21 & 2.2 & 9.0 & 3.0   & 1.3 & 4.1\\
\texttt{SLy\_14\_14}  & 0.8 & 0.41 & 1.8 & 14.6 & 3.6 & 2.4 & 11.9\\
\texttt{APR3\_13\_13} & 2.9 & 0.22 & 2.0 & 9.8 & 3.2 & 1.3 & 4.1 \\
\texttt{APR3\_14\_14} & 4.1 & 0.24 & 3.4 & 19.1 & 5.7 & 2.7 & 11.9 \\
\texttt{MPA1\_13\_13} & 1.9 & 0.21 & 1.1 & 3.4 & 1.3 & 0.5 & 0.9\\
\texttt{MPA1\_14\_14} & 3.2 & 0.23 & 2.5 & 13.7 & 4.0 & 2.1 & 8.6 \\
\texttt{MPA1\_12\_15} & 4.2 & 0.25 & 3.5 & 16.6 & 4.6 & 2.4 & 11.6 \\
\texttt{MPA1\_12\_18} & 7.0 & 0.32 & 8.1 & 29.3 & 9.0 & 3.8 & 17.6 \\
\texttt{MS1b\_13\_13} & 10.4 & 0.20 & 4.5 & 4.7 & 2.1 & 0.4 & 0.6 \\
\texttt{MS1b\_14\_14} & 12.0 & 0.21 & 6.0 & 8.3 & 3.8 & 0.5 & 0.5 \\
		\hline
	\end{tabular}
\end{table*}
%
%%%%%%%%%%%%%%%%%%%%%%%%%%%%%%%%%%%%%%%%
\begin{figure*}
\includegraphics[width=2\columnwidth]{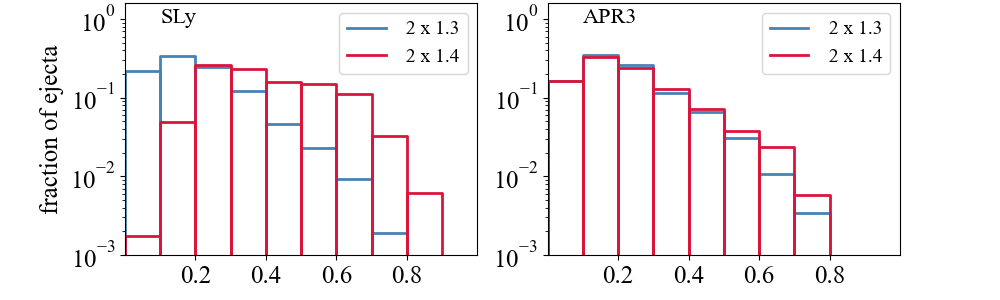}
\vspace*{0.cm}
\includegraphics[width=2\columnwidth]{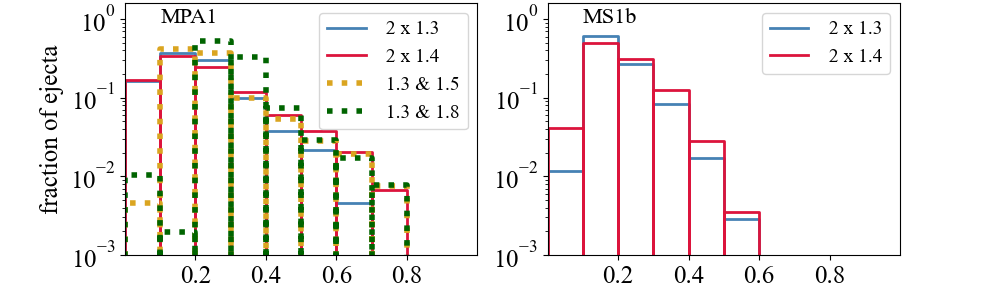}
  \caption{Ejecta fraction of {\em all ejecta} binned according to velocity at infinity, each panel shows one equation of state.}  
  \label{fig:vel_bins_all}
  \end{figure*}
%%%%%%%%%%%%%%%%%%%%%%%%%%%%%%%%%%%%%%%%
%%%%%%%%%%%%%%%%%%%%%%%%%%%%%%%%%%%%%%%%
\begin{figure*}
\includegraphics[width=2\columnwidth]{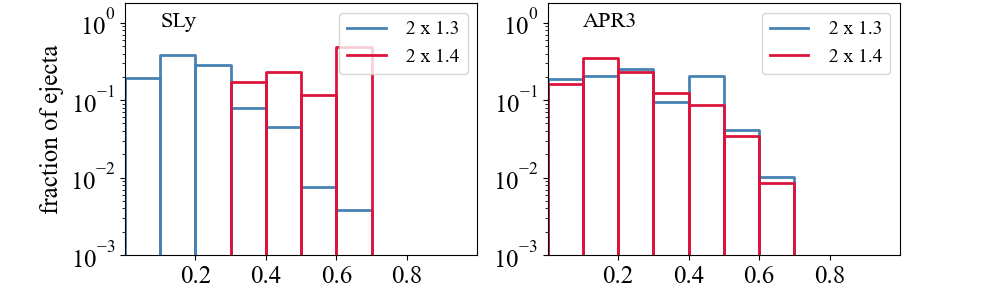}
\includegraphics[width=2\columnwidth]{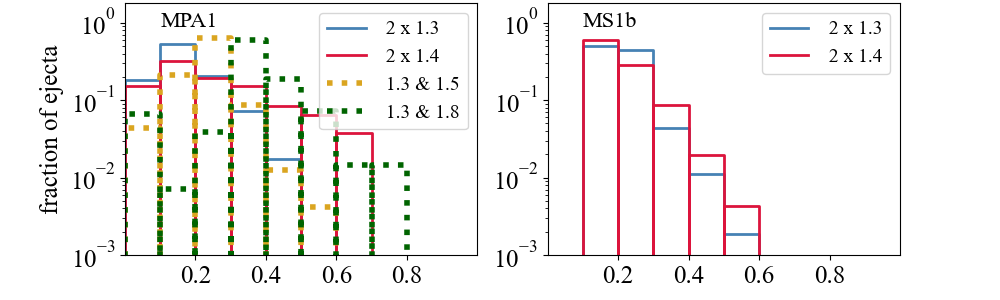}
  \caption{Ejecta fraction of the {\em polar ejecta} ($|\Theta| < 30^\circ$) binned according to velocity at infinity.} 
  \label{fig:vel_bins_pol}
\end{figure*}
%%%%%%%%%%%%%%%%%%%%%%%%%%%%%%%%%%%%%%%%
%%%%%%%%%%%%%%%%%%%%%%%%%%%%%%%%%%%%%%%%
\begin{figure*}
\includegraphics[width=2\columnwidth]{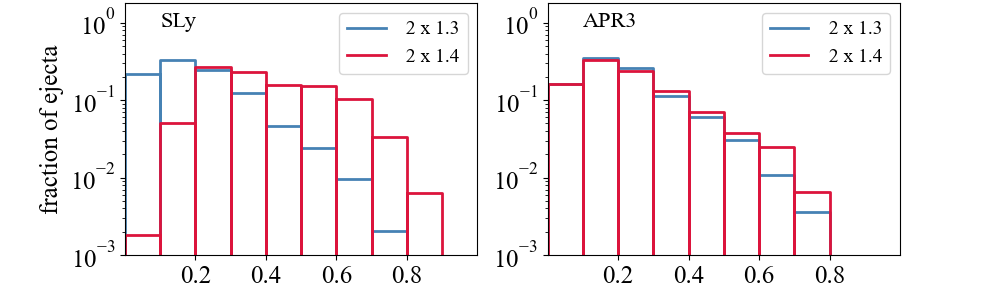}
\includegraphics[width=2\columnwidth]{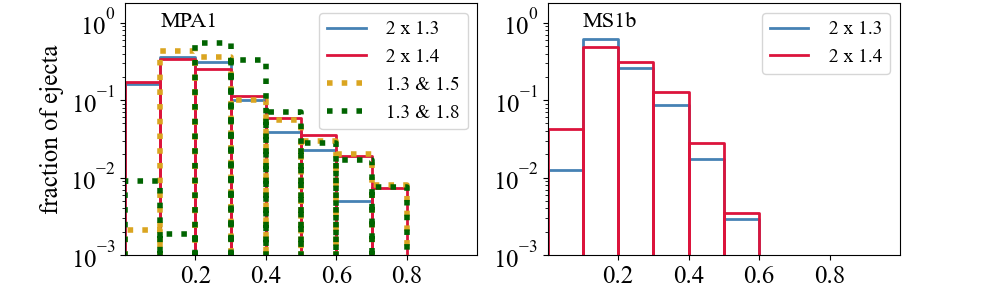}
  \caption{Ejecta fraction of the {\em equatorial ejecta} ($|\Theta| 
  \ge 30^\circ$) binned according to velocity at infinity.}  
  \label{fig:vel_bins_equ}
\end{figure*}
%%%%%%%%%%%%%%%%%%%%%%%%%%%%%%%%%%%%%%%%
Some combinations of masses and EOSs are the same as in our recent paper \citep{rosswog22b}. 
Compared to the earlier code version, we now have a slightly improved initial setup, see Sec.~3 in \cite{rosswog23a}, and a substantial improvement in our particle-to-mesh mapping, see 
Sec. 2.4 in \cite{rosswog23a}. In our earlier work we used accurate, but fixed kernel functions to perform the mapping, while now we are calculating these kernels by locally minimizing 
an error functional. With these improvements our 2 million particle runs from today look closer to our earlier 5 million particle simulations. All our runs in \cite{rosswog22b} used 
$2 \times 1.3$ \Msun and we compare to those that also used 
2 million particles. For these, we found for the SLy EOS $2.2 \times 10^{-4}$ \Msun for $v_\infty> 0.5 c$ and $8.8 \times 10^{-5}$ for $v_\infty> 0.6 c$, while 
now we find $1.3 \times 10^{-4}$ and $4.1 \times 10^{-5}$ \msun. For the APR3-EOS we found $9 \times 10^{-5}$ and $2.8 \times 10^{-5}$ while now we find 
$1.3 \times 10^{-4} $ and $4.1\times 10^{-5}$ \msun, for MPA1 we found $5.6 \times 10^{-5}$ and $1.1 \times 10^{-5}$ \msun, while the new results are 
$5.1 \times 10^{-5}$ and $9 \times 10^{-6}$ \Msun and for MS1b we found $8.8 \times 10^{-5}$ and $3.5 \times 10^{-5}$ \Msun while the new results are 
$3.6 \times 10^{-5}$ and $6.1 \times 10^{-6}$ \msun. This could be a hint that higher resolution 
could lead to smaller fast ejecta masses.
\\
We also bin the ejecta velocities according to their 
asymptotic velocity $v_\infty= \sqrt{1 - 1/(\mathcal{E} U_0)^2}$, 
see Figs.~\ref{fig:vel_bins_all} to \ref{fig:vel_bins_equ}. 
To estimate non-thermal emission, it is useful
to measure how much  mass  is faster than a given value of $\gamma \beta$, where $\gamma$ is the Lorentz factor and $\beta= v_\infty/c$. This is 
shown in Fig.~\ref{fig:M_gamma_beta}, the left panel
shows all ejecta, the right panel shows  "polar" ejecta only, defined as being within $30^\circ$ of the binary rotation 
axis\footnote{Although smaller angles would be interesting, we do not have enough particles within smaller angles to make useful statements.}.
%
%%%%%%%%%%%%%%%%%%%%%%%%%%%%%%%%%%%%%%%%

In general, for all simulations performed in this work, we find that our mass-momentum profile is well described following an exponential, $M(> \gamma\beta) \propto e^{A \gamma\beta}$, where A is in range between 2-5. This contrasts typical assumptions of a broken power-law profile based on numerical simulations~\citep{sadeh23, Zappa2023}. In particular, we find an extra shallow decay segment between $0.1 \lesssim \gamma\beta \lesssim 0.2$, after which the mass-momentum profile is more consistent with a broken power-law profile employed in previous work. Sticking with the broken power-law parameterization and ignoring the first shallow component (for ease of comparison to previous work), we fit a broken power-law profile of the form 
\begin{equation}
M(>\gamma\beta)=M_0\begin{cases}\left(\frac{\gamma\beta}{\gamma_0\beta_0}\right)^{-s_\mathrm{ft}}&\mathrm{for}\quad\gamma_0\beta_0<\gamma\beta,\\\left(\frac{\gamma\beta}{\gamma_0\beta_0}\right)^{-s_\mathrm{KN}}&\mathrm{for}\quad0.2<\gamma\beta<\gamma_0\beta_0,\end{cases}
\end{equation}
to the mass-momentum profiles for all simulations. 
Across our simulations and fitted profiles, we find that the stiffer EOSs have smaller $\gamma_0\beta_0$, with steeper mass profile (larger $s_{\mathrm{ft}}$). Meanwhile, the effect of a larger mass merger (for the same EOS) leads to a smaller $\gamma_0\beta_0$ and shallower decay (smaller $s_{\mathrm{KN}}$).\\
A noteworthy result of our simulations is the ejecta from the single case with a prompt collapse to a BH,  SLy\_14\_14. As can be seen in Fig. \ref{fig:vel_bins_all}, the ejecta in this case is dominated by the faster material. In all simulations without a prompt collapse, most of the ejecta is moving at 0.01-0.2c, with only a small fraction ($\sim 10\%$ at $v>0.4$c). However, most of the slower material is absent in the simulation with the prompt collapse since it falls into the newly formed BH. Thus, most of the ejecta move faster than 0.2c, with half of the ejecta moving faster than 0.4c. In fact, the mass of the fast ejecta ($>0.4$c) in the prompt collapse simulation, is similar to that of simulations without a prompt collapse. While we cannot 
be sure that this is a generic property of prompt collapse mergers, we see it in the one prompt collapse case of this paper, but also in simulations
with prompt collapse that are not part of this paper. This 
suggests that (at least in some cases) the observational signature of fast material in prompt collapse mergers will be similar to that of mergers that produce a longer-lived proto-neutron star. 
The primary difference between prompt collapse and longer-lived NS seen in our simulations would be in the KN emission. 
We note that here we refer to longer-lived neutron stars relative to the timescales of our numerical simulations. If the neutron star survives past the Alfven timescale, the KN afterglow will be significantly brighter due to additional energy extracted from the rotational energy reservoir of the neutron star~\citep[e.g.,][]{Sarin2022}.\\

\subsection{Kilonova afterglow}
\begin{figure}
\centerline{\includegraphics[width=1\columnwidth]{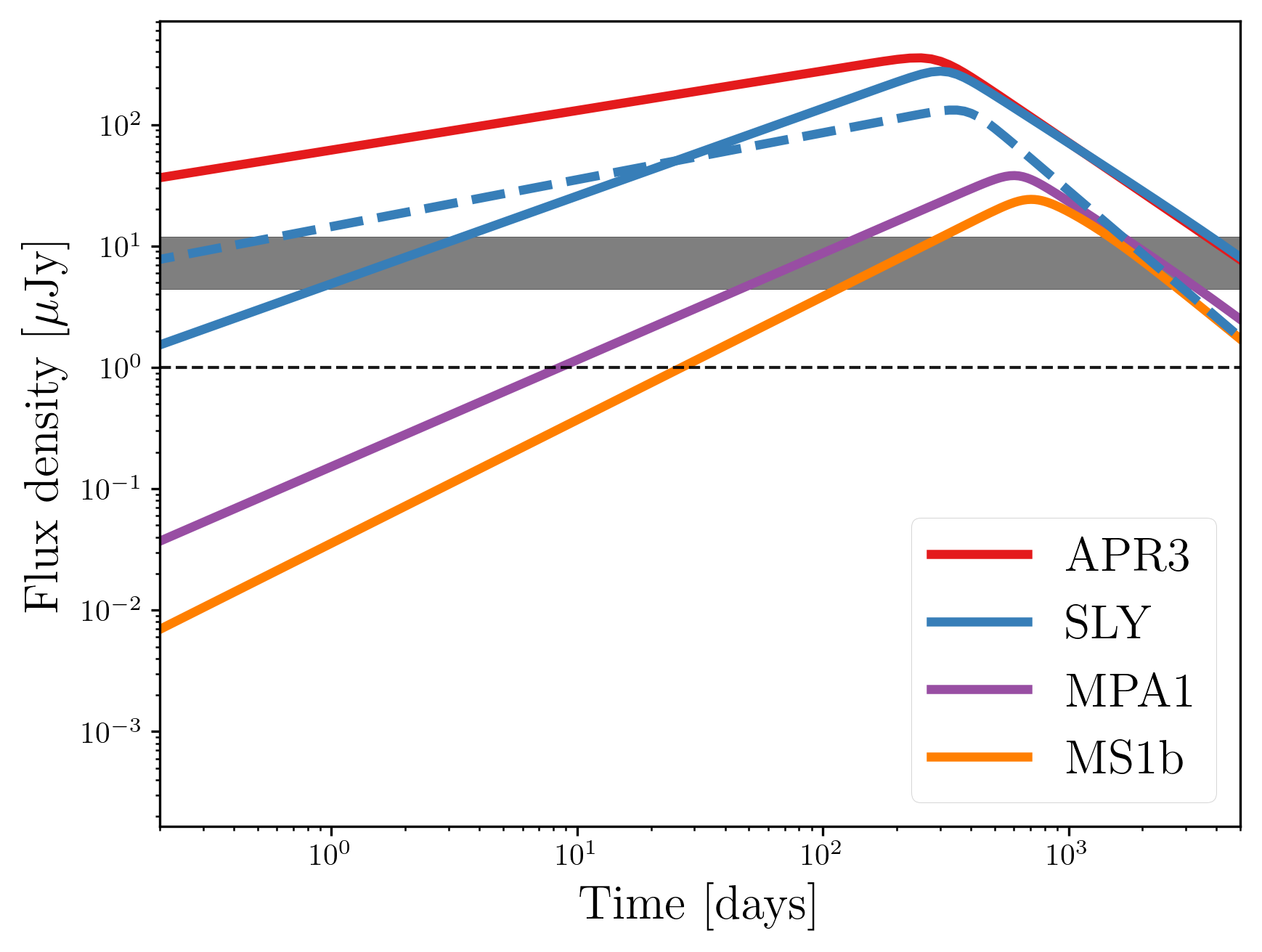}}
  \caption{Kilonova afterglow in radio (at $1$GHz) for a merger at $200$Mpc for a binary of two $1.3~M_{\odot}$ (with the prompt collapse scenario shown as a dashed blue curve) neutron stars with different EOSs. The horizontal grey band indicates the rms sensitivity limits from one hour of observing with Meerkat, while the black horizontal line indicates the one-hour continuum limit from DSA-2000. }  
  \label{fig:knafterglow}
\end{figure}

In Fig.~\ref{fig:knafterglow}, we show the kilonova afterglow in radio (at $1$GHz) for a merger at $200$ Mpc for a binary of two $1.3~M_{\odot}$ (with the prompt collapse case shown as a dashed curve) neutron stars with different EOSs following the kilonova afterglow model from \citet{sadeh23} implemented in {\sc{Redback}}~\citep{Sarin2024_redback}. In particular, we set the $\gamma_0\beta_0$, $s_{\mathrm{ft}}$, and $s_{\mathrm{KN}}$ values for each simulation as derived from our fit described above and use values consistent with other numerical work for kilonova afterglows: $M_{0} = 10^{-3}$ \msun, ambient interstellar medium density, $n=0.1$ cm$^{-3}$, electron power-law index, $p=2.5$, and microphysical parameters $\epsilon_{e}=0.1$, and $\epsilon_b = 0.01$. The horizontal grey band indicates the rms sensitivity limits from one hour of observing with Meerkat, while the black dashed line indicates the continuum sensitivity from one hour of observing with DSA-2000, highlighting that softer EOS has more potential to produce a detectable kilonova afterglow. 

An interesting result is the prediction of the kilonova afterglow for the prompt collapse scenario. In particular, the afterglow time evolution in this scenario is much more distinct compared to other simulations. 
This is a direct consequence of the mass-momentum profile, where the bulk of the material in this run is at a faster velocity than the other simulations. 
For the specific case in Fig.~\ref{fig:knafterglow}, where we have assumed a fixed $M_{0}$, the afterglow is brighter at early times as the ejecta is dominated by the faster moving material, while after peak, the lack of slower material (as it falls into the newly formed BH), results in an overall dimmer afterglow relative to SLy simulation with a longer-lived neutron star. Notably, the prompt collapse afterglow is brighter than kilonova afterglows of stiffer EOS which form longer-lived neutron stars. 
This specific impact of ejecta mass-momentum profile on the early kilonova afterglow could be used as a 'smoking-gun' and distinguish between merger outcomes. 
However, distinguishing between these different cases will likely require observations well before peak (before the deceleration timescale of the ejecta) and will likely be obfuscated by the systematic uncertainty from the uncertain microphysical parameters and ambient density medium as well as contributions from the jet afterglow.
%
%%%%%%%%%%%%%%%%%%%%%%%%%%%%%%%%%%%%%%%%
\begin{figure*}
\centerline{\includegraphics[width=2\columnwidth]{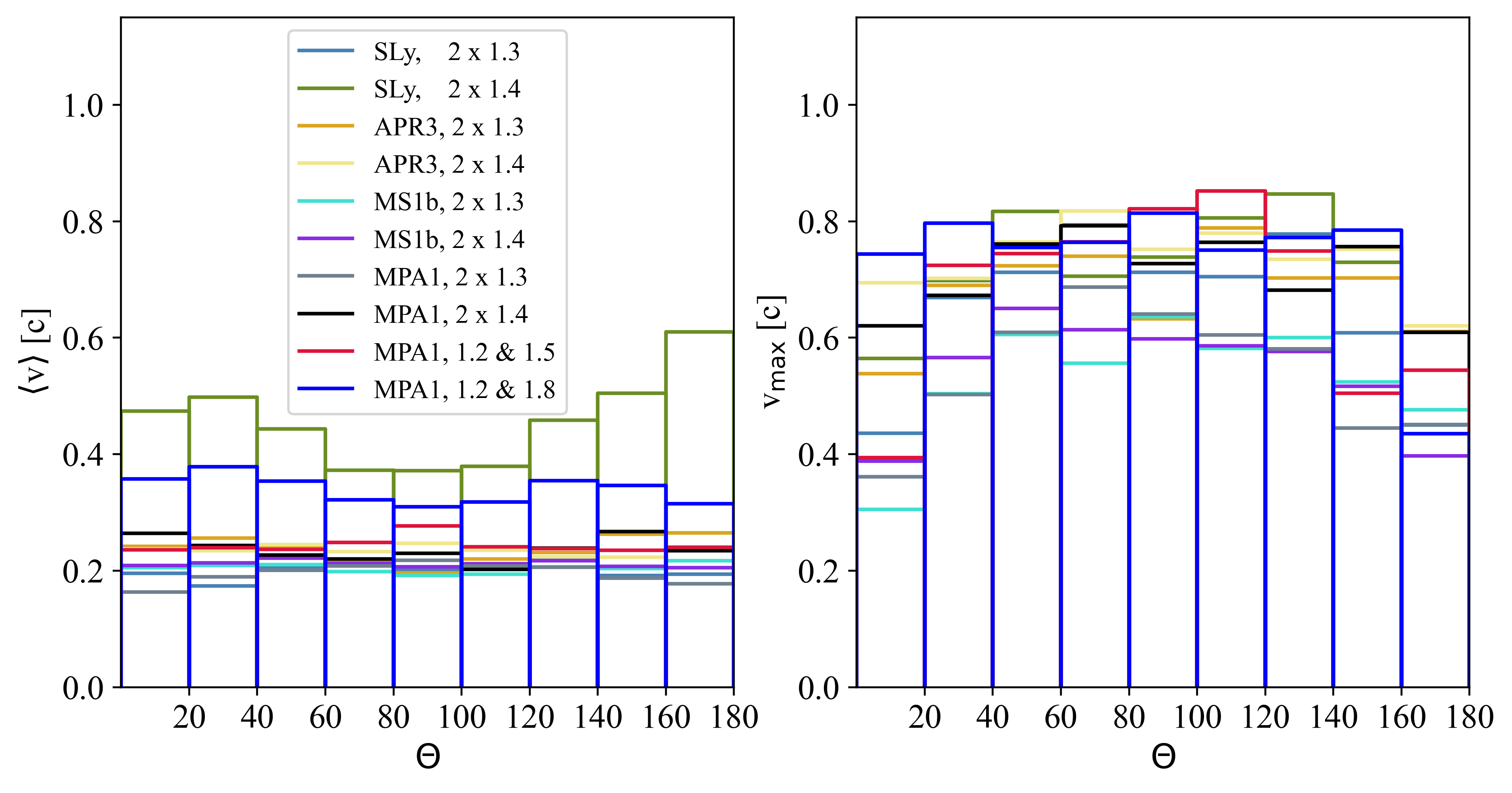}}
  \caption{Velocities as a function of the angle from the binary rotation axis (in degrees). The left panel shows the average velocity while the right panel shows the maximum velocity. All velocities are asymptotic values "at infinity".}  
  \label{fig:V_vs_theta}
\end{figure*}
%%%%%%%%%%%%%%%%%%%%%%%%%%%%%%%%%%%%%%%%

\subsection{Shock breakout}
Following the merger, at least in some cases, a relativistic jet is launched along the poles, with some delay compared to the dynamical ejecta. The propagation through the ejecta inflates a cocoon, and if the jet is powerful enough to break through the ejecta --a necessary condition for the production of a GRB-- the cocoon that it drives must break out of the fast ejecta as well. Also, if the jet is choked, the cocoon may still drive a shock strong enough to break out of the ejecta \citep[e.g.,][]{Gottlieb2018}. The breakout of the cocoon generates a flare of gamma rays, which may be detected over a much larger angle than the jet-opening angle \citep{kasliwal17,Gottlieb2018}. 
The velocity of the breakout depends on the angle from the jet axis, and it ranges from ultra- to mildly relativistic. A detailed discussion of the observed signal is provided in \cite{Nakar2020}. Here, we briefly describe the main results of this theory and apply it to the high-velocity tail of the ejecta distribution that we find in our simulations. We caution, however, that this theory is approximate to within an order of magnitude and in some regimes it is based on assumptions that still need to be validated.\\ 
The observed signal depends on three parameters: (i) the delay between the merger and the jet launching, $\Delta t$; (ii) the shock Lorentz factor (and corresponding velocity) at the time of breakout, as measured in the merger rest frame, $\gamma_{s}$ and $\beta_{\rm s}$;  and (iii) the Lorentz factor and velocity of the fast ejecta at the breakout location, $\gamma_{\rm ej,bo}$ and $\beta_{\rm ej,bo}$. Useful quantities that are based on these parameters are the shock velocity and Lorentz factor as measured in the ejecta frame at the location of the breakout (i.e., the shock velocity in the upstream frame):
\begin{equation}\label{eq:beta_stag}
    \beta_{\rm s}'=\frac{\beta_{\rm s}-\beta_{\rm ej,bo}}{1-\beta_{\rm s}\beta_{\rm ej,bo}}
\end{equation}    
and 
\begin{equation}
    \gamma_{\rm s}'=\gamma_{\rm s}\gamma_{\rm ej,bo}(1-\beta_{\rm s}\beta_{\rm ej,bo}).
\end{equation}   
With these parameters, we find the radius of the breakout, which takes place when the shock, launched roughly with the jet, gets to the breakout location in the ejecta:
\begin{equation}\label{eq:Rbo}
    R_{\rm bo} \approx c \Delta t \frac{\beta_{\rm ej,bo}}{\beta_{\rm s}-\beta_{\rm ej,bo}} \sim 6 \times 10^{10} {\rm~cm~} \Delta t_1 \gamma_{\rm ej,bo}^2, 
\end{equation}
where $\Delta t_1=\Delta t/1{~\rm s}$. Here and in the following relations the last part of the equality assumes $\gamma_{\rm ej,bo} \beta_{\rm ej,bo} \gtrsim 1$ and $\gamma_{\rm s} \gg  \gamma_{\rm ej,bo}$.
The mass that is contained within the shock transition layer at the time of breakout is:
\begin{equation}\label{eq:Mbo}
    m_{\rm bo} \approx \frac{4 \pi R_{\rm bo}^2}{\kappa\beta_{\rm s}'} \approx 4 \times 10^{-10} M_\odot ~\beta_{\rm s}'^{-1} R_{\rm bo,11}^2 \sim 2.5 \times 10^{-10} M_\odot \Delta t_1^2 \gamma_{\rm ej,bo}^4
\end{equation}
where $R_{\rm bo,11}=R_{\rm bo}/10^{11}\,$cm and the opacity $\kappa$ is taken to be $0.16$~cm$^2$~g$^{-1}$, as appropriate to fully ionized r-process ejecta. Note that $m_{\rm bo}$ is the \textit{isotropic equivalent mass} within the region seen during the breakout, namely an angle of $1/\gamma_{\rm s}$ around the line of sight. A rough estimate of the energy and duration of the breakout signal is then:
\begin{equation}
    \begin{split}
        E_{\rm bo} & \approx m_{\rm bo} c^2 \gamma_{\rm s} (\gamma_{\rm s}'-1) = 7 \times 10^{44} {\rm~erg~} \frac{\gamma_{\rm s} (\gamma_{\rm s}'-1)}{\beta_{\rm s}'} R_{\rm bo,11}^2 \\
        & \sim 4 \times 10^{44} {\rm~erg~} \Delta t_1^2 \gamma_{\rm s} \gamma_{\rm s}' \gamma_{\rm ej,bo}^4
    \end{split}
\end{equation}
and 
\begin{equation}
    t_{\rm bo} \approx \frac{R_{\rm bo}}{2 \gamma_{\rm f}^2 c}  \approx 1.6 {\rm~s~} R_{\rm bo,11} (\gamma_{\rm s} \gamma_{\rm s}')^{-2} \sim 1 {\rm~s~} \Delta t_1 \left(\frac{\gamma_{\rm ej,bo}}{\gamma_{\rm s} \gamma_{\rm s}'}\right)^{2}.
\end{equation}
where $\gamma_{\rm f}$ is the final Lorentz factor of the shocked material when the radiation is released to the observer. If $\gamma_{\rm s}' \beta_{\rm s}'<1$ then the shock does not produce pairs and $\gamma_{\rm f}=\gamma_{\rm s}$. If $\gamma_{\rm s}' \beta_{\rm s}'>1$ the shock produces pairs, and the radiation is released only after the shocked gas accelerates. A very rough approximation that fits both limits is $\gamma_{\rm f} \sim \gamma_{\rm s} \gamma_{\rm s}'$  (see discussion in \citealt{Nakar2012,Nakar2020,Faran2023}). The delay between the merger (identified by the peak of the GW emission) and the breakout signal is:
\begin{equation}
    \Delta t_{\rm GW,\gamma} \approx \Delta t + \frac{R_{\rm bo}}{2 c \gamma_{\rm s}^2} \sim  \Delta t \left(1+\left(\frac{\gamma_{\rm ej,bo}}{\gamma_{\rm s}}\right)^2\right) .
\end{equation}
If the shock is relativistic in the upstream frame (i.e., $\gamma_{\rm s}' \beta_{\rm s}' >1$) the the color temperature is 
\begin{equation}
    T_{\rm bo} \sim 50 {\rm~keV~} \gamma_{\rm f}  \sim 50 {\rm~keV~} \gamma_{\rm s} \gamma_{\rm s}' ~~~~;~~~~ \gamma_{\rm s}' \beta_{\rm s}' >1.
\end{equation}
Otherwise, it is lower and can even be in the X-rays, but its estimation is not trivial (a rough approximation is given in \citealt{Nakar2020}). 

Following the shock breakout, diffusion of radiation from the shocked gas (cooling emission) continues to contribute to the gamma-ray and X-ray signal. The first phase is named the planar phase, which continues until the shocked gas doubles its radius. The contribution from the planar phase is observed simultaneously with the breakout emission (i.e., during $t_{\rm bo}$). It can dominate over the breakout signal if the breakout mass, $m_{\rm bo}$ is much smaller than the dynamically ejected mass, $m_{\rm dyn}$, defined as the mass over which the ejecta density and velocity varies significantly. This is expected if the breakout occurs from the edge of the ejecta (namely, $v_{\rm ej,bo}$ is also the velocity of the fastest moving ejecta). In such cases,
\begin{equation}\label{eq:Epl}
        E_{\rm pl} \sim E_{\rm bo} \sqrt{\frac{m_{\rm dyn}}{m_{\rm bo}}} \approx \sqrt{m_{\rm dyn} m_{\rm bo}} c^2 \gamma_{\rm s} (\gamma_{\rm s}'-1).
\end{equation}
The spectrum of the planar phase is expected to be somewhat softer than that of the breakout signal.

From the above discussion, we see for example, that for $\gamma_{\rm s} \approx 4$, $\gamma_{\rm ej,bo}\approx 2.5$ and $\Delta t \approx  1$s, the predicted breakout signal is similar to the gamma-rays seen from GW170817 \citep{Nakar2020}, without significant contribution from the planar phase (i.e., $m_{\rm dyn} \sim {m_{\rm bo}}$). Alternatively, other consistent parameters have smaller values of  $\gamma_{\rm s}$ and $\gamma_{\rm ej,bo}$ where $E_{\rm pl}$ contributes a significant fraction of the emission, although it is less clear if in cases, where  $E_{\rm pl}$ dominates the emission, the breakout can generate the observed spectrum. 

Applying the above theory to the fast polar ejecta that we find in our simulations, we note that the maximum velocity that we see is limited by resolution. Namely, we expect ejecta which are faster than the fastest particles found in our simulations, but its mass is too low to be resolved numerically. Moreover, we do not fully trust the velocity distribution shown in Fig.~\ref{fig:M_gamma_beta} at the highest velocities, due to the small number of particles at these velocities.  Thus, we cannot predict the mass or velocity distribution of the ejecta at higher velocities than those found in the simulations. \\Therefore, we take here two limits. Case I: we assume that there is no ejecta faster than the one seen in the simulations. This is a clear lower limit. Case II: we use the power-law fit for the velocity distribution, $S_{\rm ft}$, to extrapolate the ejecta to high velocities. This is most likely an upper limit, at least for ejecta that is formed by processes that are captured by the simulations.

\begin{figure*}
\centerline{\includegraphics[width=1\columnwidth]{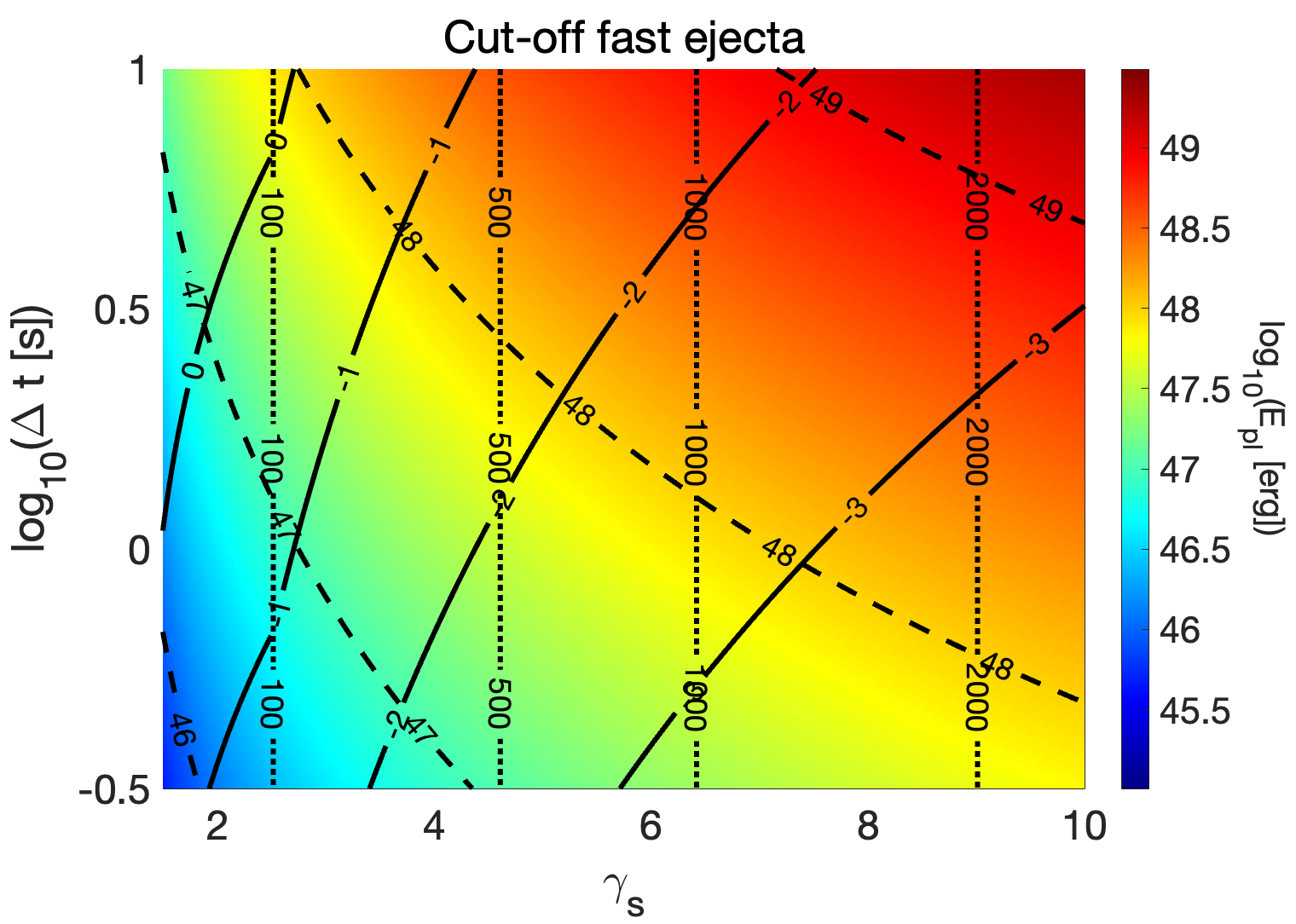}
\includegraphics[width=1\columnwidth]{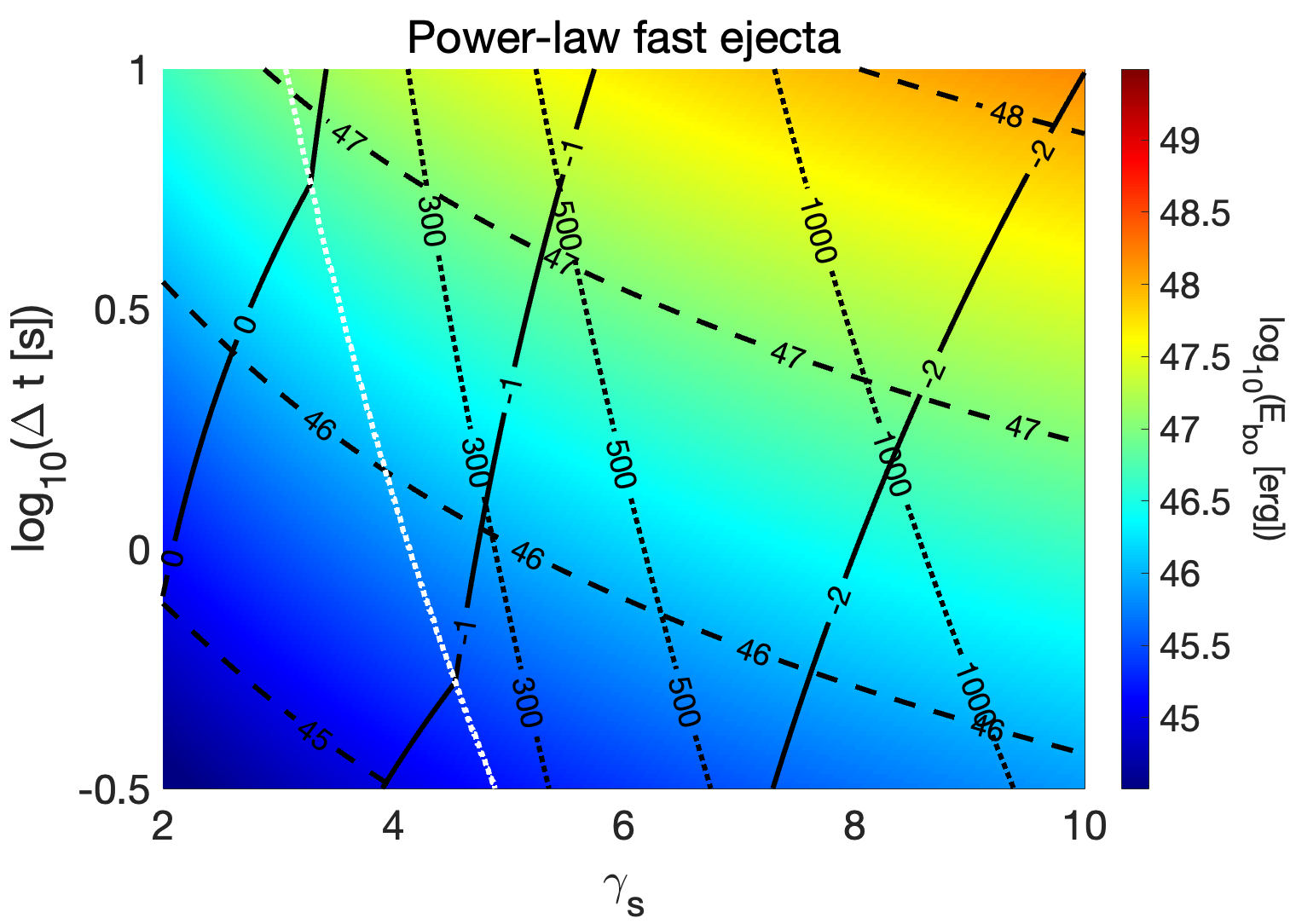}}
  \caption{Properties of the breakout emission in the phase-space of the shock breakout Lorentz factor (in the observer frame) and the time delay between the merger and the deposition of energy (e.g., in the form of a jet) that drives the shock. \textit{The accuracy of all properties is only within an order of magnitude.} Shown is the energy of the pulse in $\log_{10}([{\rm erg}])$ (colormap and dashed contours), its duration, $t_{\rm bo}$ in $\log_{10}([s])$ (solid contours), and typical photon 
  energy in keV, $t_{\rm bo}$ (dotted contours). All the properties are calculated using Eqs. \ref{eq:beta_stag}-\ref{eq:Epl}, taking into account both a relativistic and a non-relativistic shock as seen in the ejecta (upstream) frame, $\gamma_{\rm s}'$. \textit{Left panel:} \textbf{Cut-off ejecta (case I)}. Shock properties assume a sharp cut-off in the fast ejecta and no material faster than the fastest particle seen in the simulations. The ejecta parameters taken for this figure are $\beta_{\rm ej,bo}=0.6$ and $m_{\rm dyn}=10^{-5}\,M_\odot$. The energy in this plot is of the planar phase, which in this case is significantly higher than the emission from the shock transition layer. $t_{\rm bo}$ is the temperature of the radiation from the shock transition layer, which is roughly the maximal expected photon energy. The planar emission has a lower temperature. \textit{Right panel:} \textbf{A power-law ejecta (case II)}. Shock properties assume that the fast eject is extended to fast velocities, where the simulation resolution is not high enough to resolve. We assume a power-law distribution with index $\alpha=9$, $m_0=10^{-4}\,M_\odot$ and $\gamma_0\beta_0=0.4$. The white dotted contour marks  $\gamma_{\rm s}'\beta_{\rm s}'=1$. To the left of this contour, the shock is not relativistic and the observed temperature drops fast with the shock velocity, possibly even to the X-ray range. The energy and the temperature in this plot are of the photons from the shock transition layer upon breakout, which are larger or comparable to the planar phase emission.}
  \label{fig:breakout}
\end{figure*}

{\bf Case I - cut-off ejecta}: from the right panel of Fig.~\ref{fig:M_gamma_beta} we see that, for almost all simulations, around $10^{-6} M_\odot$ are moving at $\gamma\beta \approx 0.6-1$. Since this mass is confined to an angle of 30$^\circ$ from the axis, it corresponds to an isotropic equivalent mass of $\sim 10^{-5} M_\odot$. If we assume that there is no faster material along the pole than the fastest particle in that direction in the simulation, then for $R_{\rm bo} \lesssim 10^{13}$ cm ($\Delta t \lesssim 300$ s) the breakout is from the edge of the ejecta with $m_{\rm bo} \ll m_{\rm dyn} \sim 10^{-5}~M_\odot$. Taking $\beta_{\rm ej,bo}=0.6$  (corresponding to $\gamma\beta=0.75$) and  $m_{\rm dyn} = 10^{-5}~M_\odot$ as representative for case I, and assuming $\gamma_{\rm s} \gtrsim 2$ we obtain:
\begin{equation}
    R_{\rm bo} \approx  4.5 \times 10^{10} {\rm~cm~} \Delta t_1,
\end{equation}
\begin{equation}
    m_{\rm bo} \approx 10^{-10} M_\odot \Delta t_1^2,
\end{equation}
\begin{equation}
        E_{\rm bo} \sim 2 \times 10^{44} {\rm~erg~} \gamma_{\rm s} (\gamma_{\rm s}'-1) \Delta t_1^2 ,
\end{equation}
\begin{equation}
    t_{\rm bo} \approx 3 {\rm~s~} \Delta t_1 \gamma_{\rm s}^{-4},
\end{equation}
and 
\begin{equation}
    T_{\rm bo} \sim 25 {\rm~keV~} \gamma_{\rm s}^2
\end{equation}
where, in the last two equations, we used $\gamma_{\rm s}' \approx \gamma_{\rm s}/2$ as obtained for $\gamma_{\rm s} \gg 1$ and $\gamma_{\rm ej,bo}=0.6$. The delay between the GW and gamma-ray signal is 
\begin{equation}
    \Delta t_{\rm GW,\gamma} \approx \Delta t.
\end{equation}
Finally, most of the energy will be radiated by the planar phase, over a duration that is comparable to $t_{\rm bo}$:
\begin{equation}
        E_{\rm pl} \sim 3 \times 10^{46} {\rm~erg~} \gamma_{\rm s}^2 \Delta t_1.
\end{equation}
This energy is expected to be released in gamma-rays and X-rays, below $T_{\rm bo}$. 

{\bf Case II - Power-law fast ejecta}: We assume that the fast ejecta extends to high velocities as a power-law in the form 
\begin{equation}
M(>\gamma\beta)=m_0 \left(\frac{\gamma\beta}{\gamma_0\beta_0}\right)^{-\alpha}. 
\end{equation}
Based on the right panel of Fig. \ref{fig:M_gamma_beta}, we find that $m_0 \sim 10^{-4}~M_\odot$, $\gamma_0\beta_0 \approx 0.4$ and $\alpha=6-9$
provide a reasonable fit for most of our simulations (in the range of velocities in which there is enough resolution). Plugging this mass distribution to Eqs. \ref{eq:Rbo} and \ref{eq:Mbo}, assuming $\gamma_{\rm s} \gg \gamma_{\rm ej,bo}\beta_{\rm ej,bo}>1$ we obtain:
\begin{equation}
\gamma_{\rm ej,bo} \approx \left(\frac{m_0(\gamma_0\beta_0)^{\alpha} }{2.5 \times 10^{-10}~M_\odot} \Delta t_1^{-2}\right)^\frac{1}{4+\alpha} \approx 2 ~\Delta t_1^{-1/6}
\end{equation}
where the last part of the equation assumes $m_0=10^{-4}~M_\odot$, $\gamma_0\beta_0 = 0.4$ and $\alpha=8$. Note that the dependence of $\gamma_{\rm ej,bo}$ on $m_0$ and $\gamma_0\beta_0$ is weak and also the dependence on $\alpha$ in the range we find in the simulations is not very strong.  Using this value and the approximation $\gamma_{\rm s}'\approx \gamma_{\rm s}/(2\gamma_{\rm ej,bo})$ the breakout properties are:
\begin{equation}
    R_{\rm bo} \approx  2 \times 10^{11} {\rm~cm~} \Delta t_1^{2/3},
\end{equation}
\begin{equation}
    m_{\rm bo} \approx 4 \times 10^{-9} M_\odot \; \Delta t_1^{4/3},
\end{equation}
\begin{equation}
        E_{\rm bo} \sim 2 \times 10^{45} {\rm~erg~} \gamma_{\rm s}^2 \; \Delta t_1^{3/2} ,
\end{equation}
\begin{equation}
    t_{\rm bo} \sim 1 {\rm~s~} \Delta t_1^{2/3} \left(\frac{3}{\gamma_{\rm s}}\right)^{4},
\end{equation} 
and
\begin{equation}
    T_{\rm bo} \sim 50 {\rm~keV~} \Delta t_1^{1/6} \left(\frac{\gamma_{\rm s}}{4}\right)^{2},
\end{equation}
We stress that the approximation for $T_{\rm bo}$ is applicable only for $\gamma_{\rm s}'\beta_{\rm s}>1$. For weaker shocks, the temperature can be significantly lower, possibly even in the X-ray range. Finally,
\begin{equation}
    \Delta t_{\rm GW,\gamma} \approx  \Delta t \left(1+\frac{\gamma_{\rm ej,bo}^2}{\gamma_{\rm s}^2}\right) \approx  \Delta t \left(1+\frac{4 \Delta t_1^{-1/3}}{\gamma_{\rm s}^2}\right) .
\end{equation}
In this case $E_{\rm pl} \lesssim E_{\rm bo}$. 
\\
Fig.~\ref{fig:breakout} shows the values of $t_{\rm bo}$, $T_{\rm bo}$ and $E_{\rm bo}$ (case II) or $E_{\rm pl}$ (case I) in the $\gamma_{\rm s}$-$\Delta t$ phase-space. The values in this figure are calculated without assuming $\gamma_{\rm s}'\beta_{\rm s}' >1$. This figure shows that every breakout of a shock with $\gamma_{\rm s}\beta_{\rm s}>1$ generates a gamma-ray flare with energy larger than $10^{45}$ erg and a duration that is shorter than a few seconds. For $\gamma_{\rm s}<10$ and $\Delta t<10$ s the energy is smaller than $10^{49}$ erg and the duration is longer than about $0.001$ s. In all cases $\Delta t_{\rm GW} \approx \Delta t$ (note that by definition $\Delta t_{\rm GW} > \Delta t$). {Our results broadly agree with the recent calculations of \cite{Gutierrez2024}, who used several merger simulations based on Eulerian hydrodynamics simulations to estimate the cocoon breakout signal. They consider both cut-off and continuous fast ejecta (similar to our cases I \& II). For both cases, in the range breakout parameters they explore, they find results that are roughly similar to ours. }

Fig.~\ref{fig:breakout} can be compared to the characteristics of the observed gamma-ray signal of GW170817: $E \approx 4 \times 10^{46}$ erg, $t \approx 1$ s, $T \approx 100$ keV and $\Delta t_{\rm GW}=1.7$ s \citep{abbott17c}. Examining Fig.~\ref{fig:breakout} for $\Delta t \approx 1$ s (dictated by $\Delta t_{\rm GW}=1.7$ s) shows that in case I a breakout with $\gamma_{\rm s} \approx 2$ generates a flare with characteristics similar to those observed in GW 170817, while any faster shock results in a brighter flare than the one we saw. In case II, a breakout with $\gamma_{\rm s} \approx 4-5$ generates a roughly similar flare to the one observed in GW 170817. A faster breakout can be ruled out, while a slower shock produces a fainter flare. The jet that was launched in GW 170817 must generate a cocoon that drives a fast shock (at least mildly relativistic) into the fast ejecta in our direction, at an angle of  about 20$^\circ$ with respect to the jet axis \citep{mooley18,Mooley2022,GovreenSegal2023}. Our results show that this shock must produce a gamma-ray flare and that for reasonable parameters its characteristics are similar to the GRB observed in GW 170817. 

Finally a short note on the value of $\gamma_{\rm s}$. Ahead of a successful jet, it is very high and it can be comparable to the jet Lorentz factor. At larger angles, the cocoon drives a shock with a Lorentz factor that decreases with the angle, yet the breakout can be relativistic up to large angles (of order 1 rad; e.g., \citealt{Gottlieb2018}). Moreover, shocks that are strong enough accelerate even without being driven from behind if the density profile is steep enough. For example, spherical shocks accelerate in relativistic ejecta with a steep profile of the form $M(>\gamma) \propto \gamma^{-\alpha}$ if $\alpha>1$ and the shock proper velocity in the upstream frame is larger than $4/(\alpha-1)$ \citep{GovreenSegal2024}. While the shock driven by the cocoon is not spherical, its angle is wide enough to accelerate in the steep density profile we find in the simulations. This suggests, that even cocoons that are driven by choked jets may lead to a successful breakout and a significant gamma-ray emission over a relatively wide angle. 

\section{Summary and discussion}
\label{sec:summary}

\begin{figure}
	\includegraphics[width=1.\columnwidth]{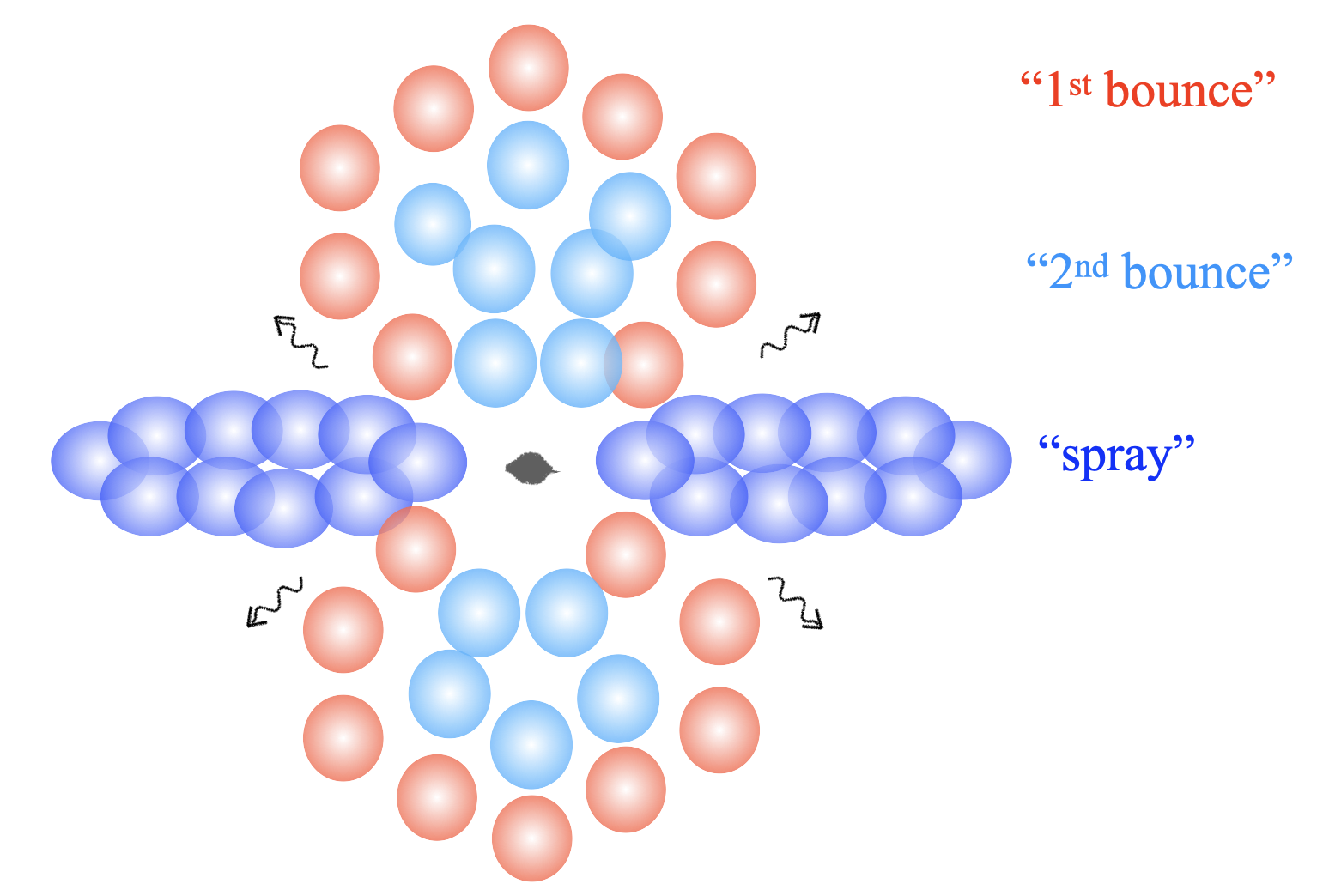}	
         \caption{Sketch of the ejection mechanisms of mildly relativistic ejecta: first matter is "sprayed" out from the interface between the two neutron stars, this matter is ejected predominantly along the orbital plane (dark blue). 
         Subsequent bounces of the central remnant launch pulses of relatively spherically ejecta that interact with the slower parts of earlier ejecta portions. This results 
         in a rather spherical shape of the first bounce ejecta, but the second bounce matter can only escape preferentially in the polar directions.}
    \label{fig:vfast_APR3_v3}
\end{figure}

In this study, we have analyzed numerical relativity simulations which were performed with the Lagrangian code \SpB for the occurrence
of fast ejecta ($>0.4c$). Interestingly, all of the studied cases eject matter with velocities of up to $\sim 0.8c$. These upper values are very likely determined by finite resolution and, in nature, higher velocities likely exist, though in amounts too small to be resolved by the current simulations. While
such fast dynamical ejecta had been seen 
in several other simulations \citep{hotokezaka13,bauswein13a,kyutoku14,kiuchi17,radice18a,hotokezaka18a,nedora21,dean21,rosswog22b,combi23}, the exact mechanism that launches these fast contributions had remained somewhat obscure.\\
Here, we find that the fast ejecta consist of two components that are launched by different mechanisms:
\bi
\i {\bf "Spray component"}:  fast matter is sprayed out at first contact from the interface between the two neutron stars, see Fig.~\ref{fig:vel_branches_APR3_evolution}, especially the second panel in the first row, and our sketch 
in Fig.~\ref{fig:vfast_APR3_v3}. This matter remains close to the orbital plane
and reaches (at our resolution) up to
$\sim 0.7c$. This component typically constitutes $\sim 30$ \% of the ejecta with $v>0.4c$.
\i {\bf "Bounce component"}: this matter is also sprayed out at first contact, but it remains initially close the central remnant. The latter becomes deeply compressed as a result of the collision, and, on bouncing back, the central remnant accelerates this matter
to  $\sim 0.8c$ (at our resolution). This
matter is ejected more spherically, see Fig.~\ref{fig:vel_branches_APR3_evolution}
and our sketch Fig.~\ref{fig:vfast_APR3_v3}. Since it is often faster than the spray component, it can catch up with slower parts of the spray component which may result in radiation. Depending on the lifetime of the central remnant, several subsequent but continuously weakening pulses may be launched. 
\ei
The fastest ejecta are initially located in an equatorial belt around the orbital plane, see Fig.~\ref{fig:vel_bins_APR3_evolution},
while slower ejecta are spread more evenly over the stellar surface. The whole
ejection period of the fast ejecta  typically lasts for only $\sim 2$ ms.

The presence and properties of the fast ejecta has a marked impact on the kilonova afterglow. Across our simulations, we see a departure from the assumptions built upon previous numerical simulations that $M(> \gamma\beta)$ follows a broken power-law profile~\citep[e.g.,][]{sadeh23,Zappa2023}, finding an extra shallow decay segment between $0.1 \lesssim \gamma\beta \lesssim 0.2$, or a profile more adequately described following an exponential. We find in general that stiffer EOS have steeper mass profiles, while larger total mass mergers (for the same EOS) have a shallower initial decay. We make predictions for the kilonova afterglow for a subset of our simulations (including the run with a prompt collapse), finding that at a fiducial distance of 200 Mpc, the radio afterglow will be observable well before peak for the softer EOS with Meerkat and for all simulations with an instrument such as DSA-2000. 

Notably, we find that even the simulation with a prompt collapse produces a bright kilonova afterglow comparable to simulations with the same EOS (SLy) that make a longer-lived neutron star. We also find that the difference in mass-momentum profile between the prompt collapse vs longer-lived NS case is prevalent before the light curve peaks. This is a direct consequence of the ejecta in the prompt scenario being dominated by ejecta at velocities $\gtrsim 0.4$c as opposed to the simulations with a longer-lived neutron star where the bulk of the ejecta is travelling at $0.2$c.
For our choice of fiducial parameters, the impact of the difference in mass-momentum profile between the prompt collapse and longer-lived neutron star simulations is largest at $t \lesssim 10$ d post-merger, providing a reason to prioritize early time radio follow-up of binary mergers. However, we caution against using such observations to determine the fate of a merger, at this timescale, we are likely to suffer significantly from systematic uncertainty in afterglow modelling, as well as contributions from the relativistic jet (apart from scenarios where we observe the system well off-axis). 
\\
The launching of a jet following the merger inflates a cocoon in the ejecta. The cocoon drives a shock that breaks out of the fast ejecta, generating a flare of gamma rays, which may be detected over a much larger angle than the jet-opening angle \citep{kasliwal17,Gottlieb2018}. The properties of this flare depend on the fast ejecta and here we use the simulations' results to estimate the breakout emission. Since the breakout depends on a tiny fraction of the ejecta mass, orders of magnitude lower than any simulation can resolve, we considered two options. In one we assume that there are no faster ejecta than what we see in the simulations and in the other that there is a power-law distribution of fast ejecta, extending to velocities higher than those that are found in the simulation. Although the predicted signal differs somewhat between these two options, in both cases the prediction is a short flare of gamma-rays that depend mostly on the shock Lorentz factor upon breakout. For mildly relativistic shock breakout the prediction is a $\sim 100$ keV flare with $\sim 10^{45}-10^{47}$ erg and a duration of $\sim 0.1-1$ s. A breakout of a highly relativistic shock with $\gamma_{\rm s} \sim 10$ produces a $\sim$ MeV flare with $\sim 10^{46}-10^{49}$ erg and a duration of $\sim$ 0.001-0.01 s. In both limiting cases, we find a region in the phase space that agrees relatively well with the observed properties of the GRB from GW 170817. In both cases, the required time-delay between the merger and the jet launching is about 1 s. The shock Lorentz factor is $\gamma_{\rm s} \approx 2$ if there is no ejecta faster than 0.6 c and $\gamma_{\rm s} \approx 4-5$ if the fast ejecta distribution is extended as a power-law to mildly relativistic velocities.\\
Clearly, this important topic has not yet been settled. Although we think that the identified ejection mechanisms are robust, we are still far from converged numbers for these small amounts of matter in fully general relativistic simulations. Changes in numerical methodology, pure numerical resolution, and also (the approximation of) different physical processes may still impact the exact amount of fast ejecta. 
Such physical processes certainly include thermal effects \citep{bauswein13a} and magnetic fields \citep{kiuchi24,aguilera24}, but neutrino physics was also found by \cite{radice16a} to affect the {\em dynamic} (and not just the secular) ejecta. Further explorations of these topics are left to future studies.

\section*{Acknowledgements}
SR would like to thank Eli Waxman for the invitation to a very inspiring
workshop on the GW-EM connection at the Weizmann Institute. Thanks also to all the
participants for creating a constructive brainstorming atmosphere.
SR has been supported by the Swedish Research Council (VR) under 
grant number 2020-05044 and by the research environment grant
``Gravitational Radiation and Electromagnetic Astrophysical
Transients'' (GREAT) funded by the Swedish Research Council (VR) 
under Dnr 2016-06012, by the Knut and Alice Wallenberg Foundation
under grant Dnr. KAW 2019.0112, by Deutsche Forschungsgemeinschaft 
(DFG, German Research Foundation) under Gemany's Excellence Strategy 
- EXC 2121 "Quantum Universe" - 390833306 and by the European Research 
Council (ERC) Advanced Grant INSPIRATION under the European Union's 
Horizon 2020 research and innovation programme (Grant agreement No. 
101053985).
N. S. acknowledges support from the Knut and Alice Wallenberg Foundation 
through the "Gravity Meets Light" project. EN was partially supported by 
a consolidator ERC grant 818899 (JetNS) and an ISF grant (1995/21). SR's calculations were performed in part
 at the NHR Center NHR@ZIB, jointly supported by the Federal Ministry of Education and Research and the state governments participating in the NHR (www.nhr-verein.de/unsere-partner), at the SUNRISE HPC 
facility supported by the Technical Division at the Department of 
Physics, Stockholm University, 
and on the HUMMEL2 cluster funded 
 by the Deutsche Forschungsgemeinschaft  (498394658). Special thanks 
 go to Mikica Kocic (SU), Thomas Orgis and Hinnerk St\"uben (both UHH) for their 
 excellent support.

%%%%%%%%%%%%%%%%%%%%%%%%%%%%%%%%%%%%%%%%%%%%%%%%%%
\section*{Data Availability}
The data underlying this article will be shared on reasonable request to the corresponding author.



\bibliographystyle{mnras}
\bibliography{astro_SKR.bib, example.bib} % if your bibtex file is called example.bib

% Don't change these lines
\bsp	% typesetting comment
\label{lastpage}
\end{document}